\documentclass[12pt]{spieman}  
\usepackage{amsmath,amsfonts,amssymb}
\usepackage{graphicx}
\usepackage{setspace}
\usepackage{tocloft}
\usepackage{lineno}
\usepackage{tabularx}
\usepackage[printonlyused,nohyperlinks]{acronym}

\graphicspath{{./}{figures/}}
\title{Habitable Worlds Observatory's Concept and Technology Maturation: Initial Feasibility and Trade Space Exploration}

\author[a,*]{Lee D. Feinberg}
\author[a]{Breann N. Sitarski}
\author[a]{Michael W. McElwain}
\author[a]{Giada N. Arney}
\author[b]{Caleb Baker}
\author[a]{Matthew R. Bolcar}
\author[b]{Marie Levine}
\author[a]{Alice Liu}
\author[b]{Bertrand Mennesson}
\author[a]{Aki Roberge}
\author[a]{J. Scott Smith}
\author[b]{Feng Zhao}
\author[b]{John Ziemer}
\affil[a]{NASA Goddard Space Flight Center, Greenbelt, Maryland 20771}
\affil[b]{Jet Propulsion Laboratory, Pasadena, CA 91109}

\cftpagenumbersoff{figure}
\cftpagenumbersoff{table} 
\begin{document} 
\maketitle

\begin{abstract}
The \ac{hwo} is the first telescope ever designed to search for life and will be a powerhouse of discovery across topics in astrophysics.  The observatory was the top recommendation of the Astro2020 Decadal Survey for large missions and a new HWO Technology Maturation Project Office was formed in August 2024 to mature the architecture, science and technology.  In this paper we review the overall approach taken to mature the mission concept.  We show progress on architecture development, integrated modeling, science cases, and technology roadmaps consistent with pre-formulation studies.  We discuss plans for instrument studies and international engagement and science engagement including a Community Science and Instrument Team.  Finally, we describe the plan forward to the Mission Concept Review. 
\end{abstract}


{\noindent \footnotesize\textbf{*}Lee Feinberg,  \linkable{lee.d.feinberg@nasa.gov} }

\begin{spacing}{2} 

\section{Introduction}
\label{sect:intro}  
The 2020 Decadal Survey on Astronomy and Astrophysics\cite{Astro2020} \ac{astro2020} recommended that NASA begin development of a large ultraviolet, optical, infrared telescope to search for signs of life and to perform transformative astrophysics.  The recommendation included a maturation of the technology, architecture and science followed by an independent review prior to the \ac{mcr}.  The effort was bolstered considerably on August 1$^{st}$, 2025 when a new project office was formed for the \ac{hwo}, led by NASA's \ac{gsfc} in collaboration with the \ac{jpl} to advance upon earlier efforts to study the science and technology(see Section~\ref{sec:science}). The new project office has been working to address the entire concept including the architecture, technology and science through use of \ac{cml} framework\cite{Wessen2013} and this paper demonstrates the advancement of the entire concept and described the plan forward toward \ac{mcr}. 

This paper presents the maturation progress of \ac{hwo} during \ac{cml}2-3, following the mission architecture development approach recommended by the 2020 Decadal Survey. Section 2 describes the overall approach for concept maturation through iterative Exploratory Analytic Cases, including the evaluation of architecture trade space and launch vehicle considerations. Section 3 presents the baseline \ac{eac} configurations studied between \ac{cml}~2 and \ac{cml}~3 the refined \ac{eac}~4 and \ac{eac}~5 designs to be developed and assessed between \ac{cml}~3 and \ac{cml}~4. Section 4 details the integrated modeling methodology and results, demonstrating the multi-thread modeling pipeline approach used to assess end-to-end Observatory performance. Section 5 summarizes the technology development program, including the prioritization framework for critical, urgent, and long-term technologies required to achieve the goal of \ac{trl}~5 prior to \ac{mcr}. Section 6 describes the testbed facilities and infrastructure supporting technology maturation activities. Section 7 addresses key engineering challenges including ultra-stable structures, wavefront sensing and control, and metrology requirements. Finally, Section 8 presents a summary of the early science case exploration developed through community engagement with the \ac{start}, and being further advanced with the \ac{csit}, covering the wide range of science enabled by this transformative observatory.

\section{Pioneering Approach for Flagship Concept Maturation} 

Our novel flagship maturation approach follows the recommendations of \ac{astro2020}, the Large Missions Study\cite{LargeMissionStudy}, and the \ac{jwst} Lessons Learned\cite{JWST_LL} that all emphasized the importance of maturing the architecture, technology, and science iteratively and holistically from the early concepts. This approach to flagship missions is consistent with the best practices for carrying out a wide range megaprojects, which encourages projects to ``think slow, act fast"\cite{Flyvbjerg2023}. Once a detailed plan is in place at the \ac{mcr}, the project will be ready to ramp up and carry out development on a fixed schedule at a rapid pace.  Early maturation of the \ac{ngrst} architecture has enabled the team to execute its development on cost and schedule using elements of this approach; however, the \ac{ngrst} was built around a legacy telescope, which is the primary architectural decision for a flagship observatory. \ac{htmpo} is using the \ac{cml} framework to explore a wide range of telescope and observatory architectures while maturing the entire concept holistically. 

\begin{figure}
\begin{center}
\begin{tabular}{c}
\includegraphics[height=5.3cm]{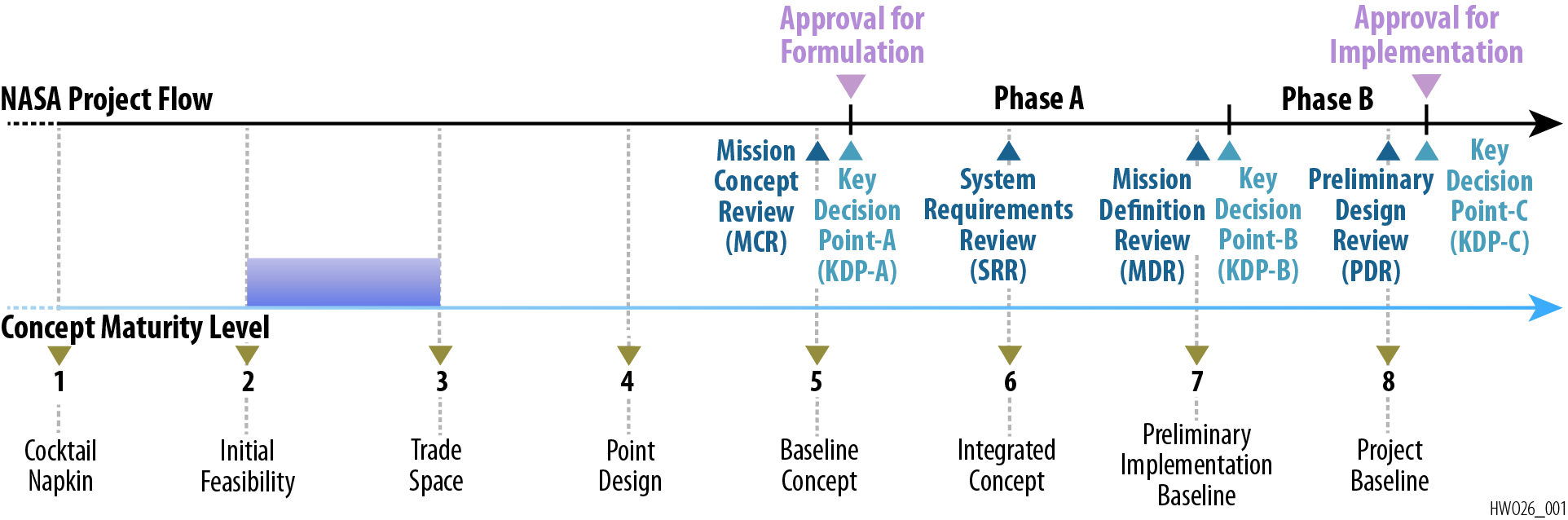}
\end{tabular}
\end{center}
\caption 
{ \label{fig:NASA+CMLs}
The traditional NASA flight project requirements include a series of independent reviews and key decision points. The \ac{cml} flow parallels in the NASA flight project flow but adds criteria for additional dimensions of concept development. The \ac{hwo} phase reported herein represents NASA pre-formulation and covers \ac{cml}~2-3, shaded purple on this timeline.} 
\end{figure}

The \ac{cml} approach starts with an exploratory phase of initial feasibility transitioning to a comprehensive evaluation of the trade space. As illustrated in Figure \ref{fig:NASA+CMLs}, between \ac{cml}~2 and \ac{cml}~3, the high-level system concept is explored by opening the trade space, framing key questions, and analyzing the driving needs. At \ac{cml}~3, the trade space is understood.  Between \ac{cml}~3 and \ac{cml}~4, there is a value framework established, with an assessment of the trade space, and a prioritization of paths forward.  To mature the engineering through this exploratory phase, we have used the \ac{eac} architectures and their integrated models to assess their relative science performance. These architectures are accompanied by science cases and a technology plan thus addressing the critical areas of \ac{cml}~3.

The \ac{cml} approach includes six concept dimensions to mature: story, implementation, science, cost, strategy, and engineering\cite{Wessen2013}.  This approach has been adopted for \ac{hwo} where the science, technology, and architecture are focusing on these ares of development while pioneering this for a large flagship mission. Per this approach, at this early \ac{cml} phase the team focuses on complexity and cost risk in addition to the strategy for completing the entrance criteria for the \ac{mcr} and the establishment of a Pre-Formulation Plan. At \ac{cml}~5 the entire mission concept will be sufficiently mature to convene the \ac{mcr} and proceed into the formulation phase.    

\section{Architectural Exploration of the Trade Space} %
Exploratory Analytic Cases (EACs) serve as preliminary investigations of initial architecture options, with configurations strategically selected to probe critical parameter space at the architectural level. These \ac{eac}s are initially developed at lower fidelity, as they are not intended to replicate actual flight designs but rather to facilitate end-to-end modeling integration from science requirements through engineering implementation. The \ac{eac} approach enables the development and validation of initial models and computational frameworks through representative examples that ``pipe-clean'' analytical processes, assess end-to-end modeling capabilities and requirements, identify critical technology gaps, and guide the maturation of potential technology solutions. Furthermore, \ac{eac}s explore key architectural trade spaces and decision breakpoints within the context of launch vehicle constraints, thereby informing future point design selections and providing timely input to launch vehicle vendors who can consider their development plans with this in mind.

The \ac{eac} designs are not expected to not expected to be the final design of \ac{hwo}. It is expected that through a rapid iterative process the design will be better understood and matured. Later iterations will be increasingly detailed. A similar evolution occurred during \ac{jwst} development as can be seen in Figure~\ref{fig:JWSTComp} below. In the case of \ac{jwst}, we compare the yardstick design with its final design\cite{McElwain2023}. The \ac{hwo} \ac{eac}s represent an earlier iteration than the \ac{jwst} yardstick.  This reinforces that the \ac{eac} work is high-level architecture evaluation and should not be overly interpreted as design.

\begin{figure}
\begin{center}
\begin{tabular}{c}
\includegraphics[height=6cm]{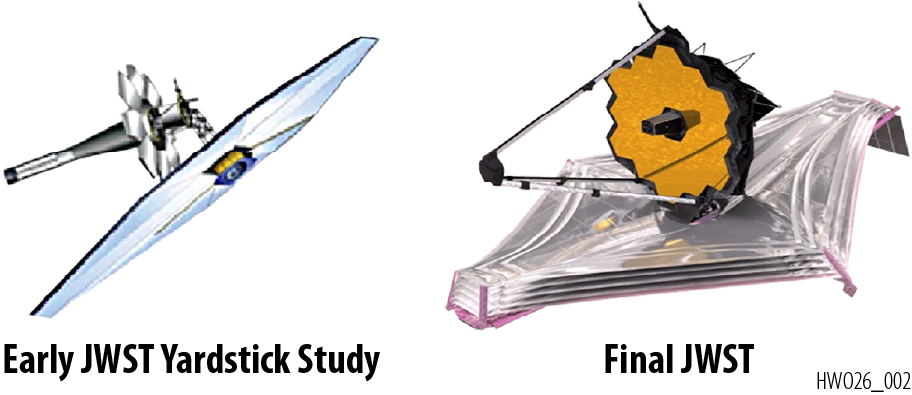}
\end{tabular}
\end{center}
\caption 
{ \label{fig:JWSTComp}
The early \ac{jwst} yardstick study was an early study that demonstrated the feasibility of the architecture. The \ac{jwst} yardstick included general architecture elements that were included in the final \ac{jwst} design. The \ac{hwo} \ac{eac} studies are even less mature than the \ac{jwst} yardstick.} 
\end{figure} 

\begin{figure}
\begin{center}
\begin{tabular}{c}
\includegraphics[height= 4.8cm]{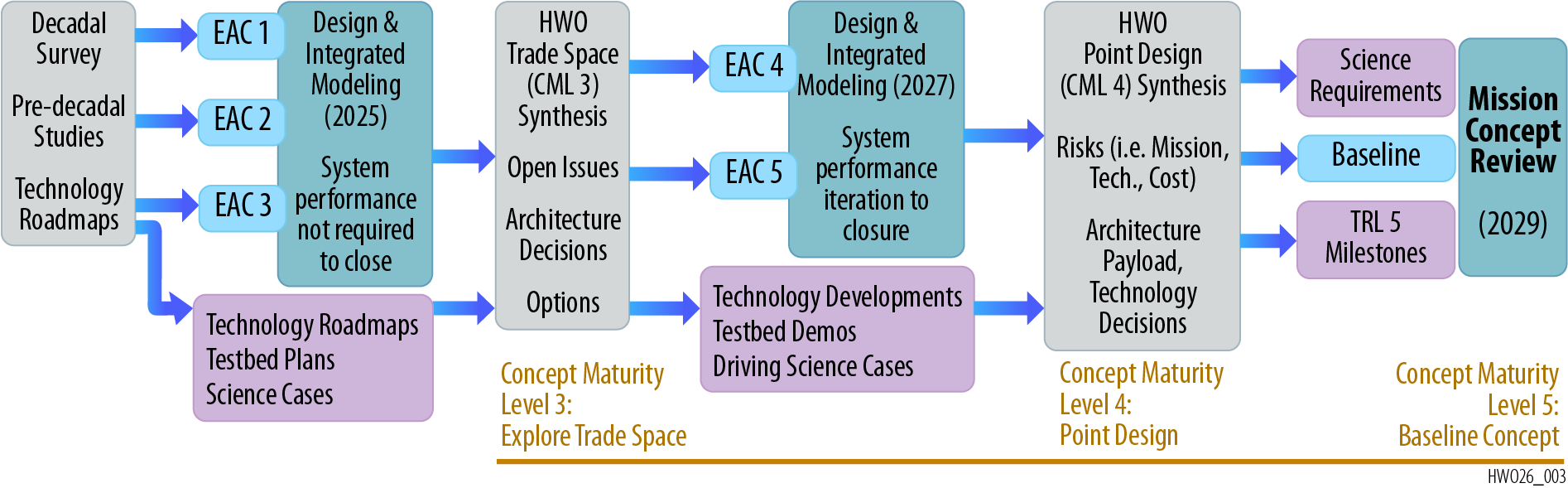}
\end{tabular}
\end{center}
\caption 
{ \label{fig:EACFlow}
The \ac{hwo} project has explored the observatory trade space by carrying out iterative architectural designs, each called an Exploratory Analytic Case (EAC). Each design is evaluated for science performance and technical feasibility, providing important insights for future iterations. This report includes material developed through \ac{cml}~3.} 
\end{figure}

The basic approach to \ac{eac}s is to iterate as we advance \ac{cml} levels. \ac{eac}-1, -2 and -3 were explored at a high level leading up to \ac{cml}3, \ac{eac}-4 and -5 will be studied at a more detailed level for \ac{cml}4, and a point design will be studied between \ac{cml}~4-5 that will demonstrate basic feasibility of the architecture.  Finally, at \ac{cml}~5 there will be a single point design consistent with mission objectives which can be taken to the Mission Concept Review.  The basic flow of the \ac{eac}s and how they iterate with technology and science can be seen in Figure~\ref{fig:EACFlow}.

\begin{figure}
\begin{center}
\begin{tabular}{c}
\includegraphics[height=7cm]{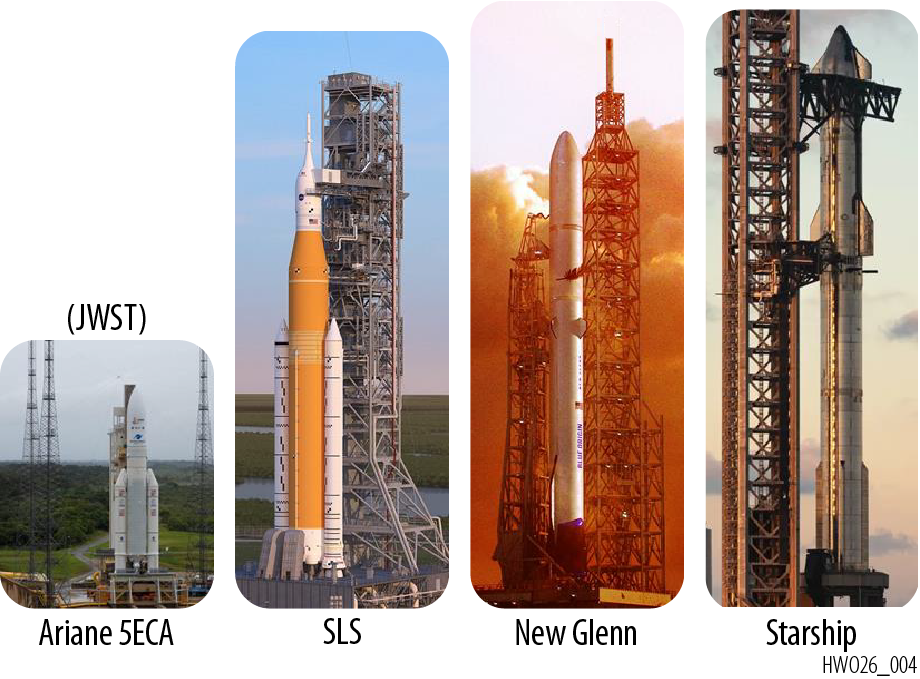}
\end{tabular}
\end{center}
\caption 
{ \label{fig:Launchers}
The launch vehicles under consideration for \ac{hwo} have considerably more mass and volume capabilities compared to the previous generation. Relative-scale images of the Arianespace Ariane 5, which launched \ac{jwst}, next to the NASA \ac{sls}, Blue Origin New Glenn, and SpaceX Starship.} 
\end{figure}

Another key consideration is the next generation of launch vehicles. At this early point, there are three key large launch vehicles being tracked by \ac{htmpo} (see Figure~\ref{fig:Launchers}).  NASA's \ac{sls} \cite{SLSUsersGuide} was designed to carry a crewed Artemis program mission to the moon. While the \ac{sls} has successfully launched and has good mass capability, a large fairing is not currently in development.  The SpaceX Starship \cite{StarshipUsersGuide} has a very large fairing and excellent mass capability but it still needs to demonstrate full reuseability, as the \ac{hwo} would require refueling in space to get to Sun-Earth L2.  Finally, the first-generation Blue Origin New Glenn \cite{NewGlenn} has launched and provides modest mass and volume improvements over current state of the art, but a next generation Blue Origin launch vehicle could provide a comparable fairing volume to the Starship. It is important for \ac{htmpo} to explore architectures that are flexible to accommodate the uncertain futures of launch vehicles. Some architectural flexibility can be gained by having a segmented primary mirror that can be folded for launch, a deployable secondary mirror structure, and a modular design that is capable of in-space integration.  

\begin{figure}
\begin{center}
\begin{tabular}{c}
\includegraphics[height=7cm]{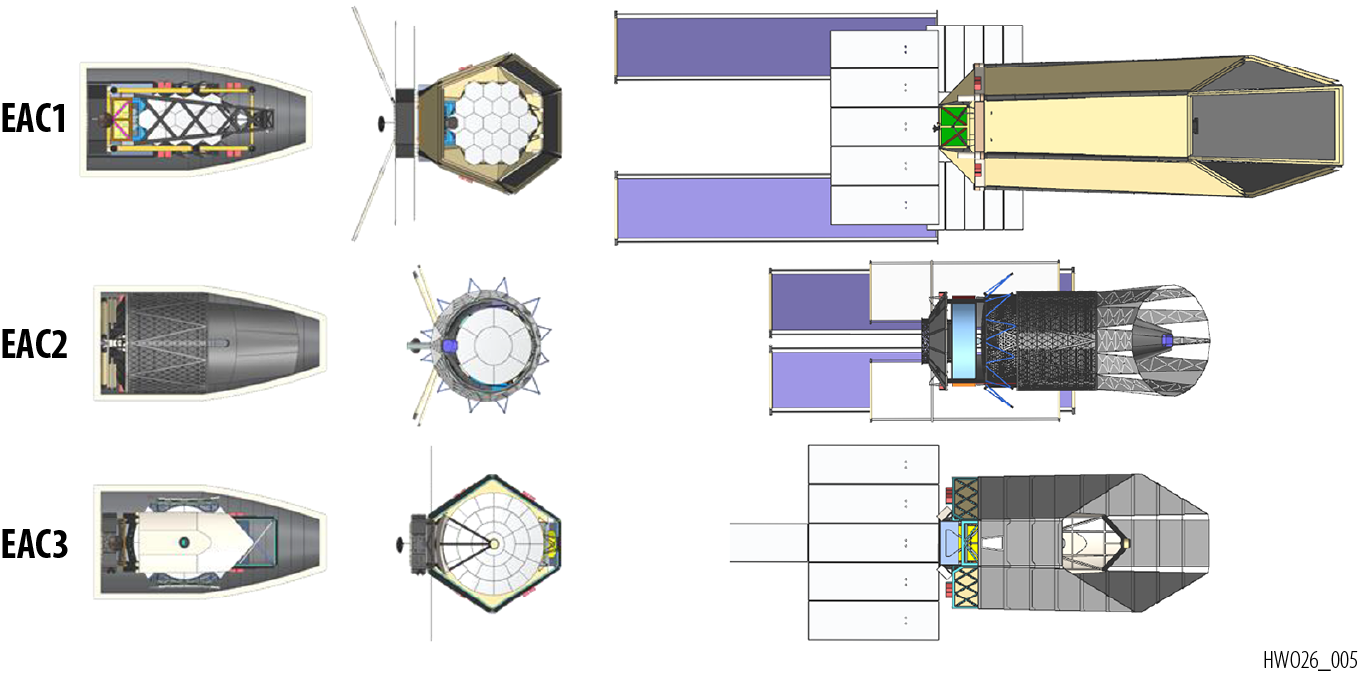}
\end{tabular}
\end{center}
\caption 
{ \label{fig:EACs123}
The \ac{eac}-1 (top row), \ac{eac}-2 (middle row), and \ac{eac}-3 (bottom row) designs shown to relative scale. The left column shows the stowed, launch configuration for each architecture and its fit inside the representative Starship fairing. The middle column is looking down the telescope boresight, and the right column shows the top-down view. A series of well-established deployments are used to transform between the stowed and operational configurations.} 
\end{figure}

To study the \ac{eac}s and probe parameter space including consideration of future large rockets, our team defined three \ac{eac}s shown in Figure~\ref{fig:EACs123}.  \ac{eac}~1 is an off-axis telescope architecture (secondary mirror does not obscure the aperture) with a 6~m inscribed primary mirror diameter and a 7.2~m outer diameter using hexagonal segments.  \ac{eac}~2 has a 6~m circular aperture using keystone segments and is also off-axis.  \ac{eac}~3 is on-axis and 8~m outer diameter with keystone segments. All studies have assumed \ac{ule} glass though other materials such as Zerodur will be considered. The configuration parameters for \ac{eac}1-3 are given in Table \ref{tab:eac123_tab}. 

\begin{figure}
\begin{center}
\begin{tabular}{c}
\includegraphics[height=5.2cm]{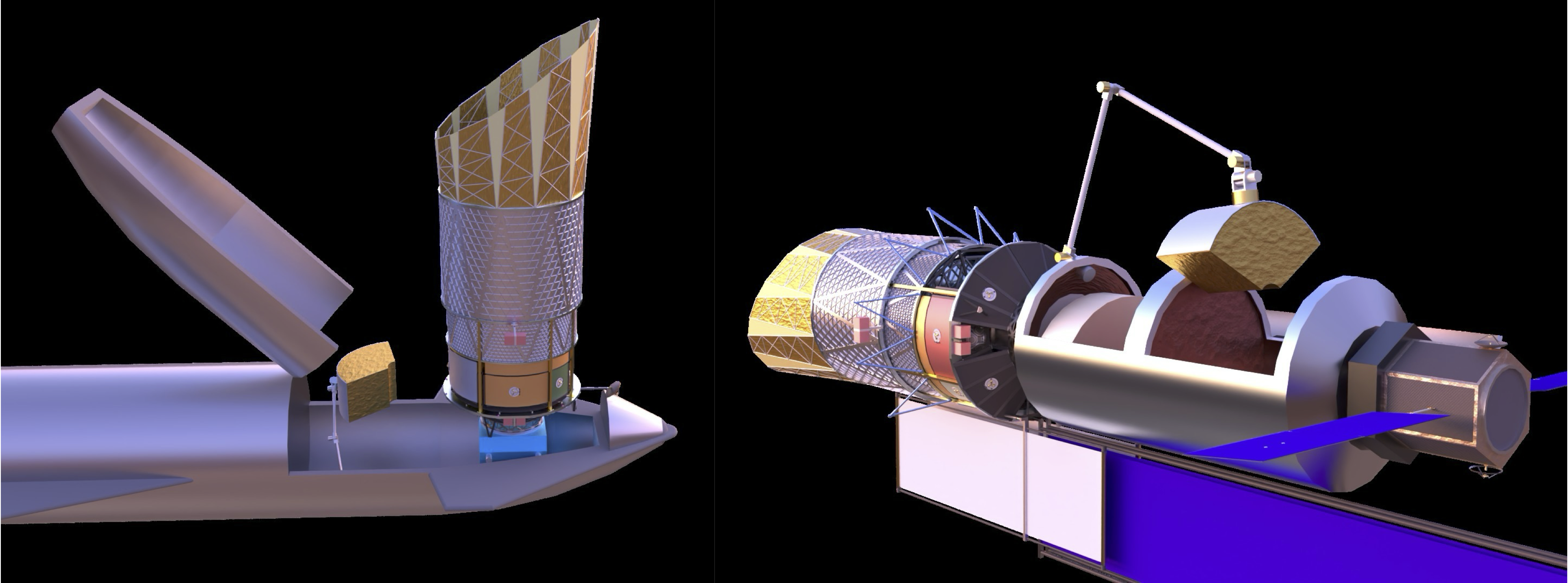}
\end{tabular}
\end{center}
\caption{
\label{fig:servicing}
Two different servicing concepts for the EACs. The left panel shows the Hubble-shuttle model, while the right shows a dedicated servicer. 
}
\end{figure}

The \ac{hwo} architecture for \ac{mcr} will incorporate on-orbit serviceability as a key design consideration. The observatory is being designed to include a modular systems architecture with standardized mechanical, thermal, electrical, and data interfaces conforming to orbital replacement unit design principles, which also streamlines ground-based integration and test activities. For a servicing mission, \ac{hwo} would use capabilities that are now being developed for geosynchronous satellites and Moon to Mars concepts.  Various mission ``concept of operations" allows for servicing scenarios utilizing emerging commercial robotic servicing capabilities such as shown in Figure~\ref{fig:servicing}. Near-term industry development efforts include generating servicing mission architectural concepts, as part of the project’s D.19 HWO System Technology Demonstrations and Mission Architecture Studies program, that can be used to identify engineering gaps. These will inform and motivate industry to be prepared for new opportunities requiring more autonomy, guidance, navigation and control measures optimized at orbitable distances beyond low-Earth and geosynchronous orbits. The resulting architecture could be used to optimize lifecycle cost-performance metrics by distributing technology development risk across multiple servicing opportunities, accommodating technology infusion as capabilities mature, and extending mission lifetime through select system replacement and augmentation analogous to Hubble servicing missions. 

\begin{table}
    \centering
    \begin{tabularx}{\textwidth}{X X X X}\hline
         \textbf{Configuration Parameter}&  \textbf{EAC1}& \textbf{EAC2} & \textbf{EAC3} \\\hline
         Launch Vehicle Compatibility&  Blue Origin's New Glenn or Starship (7 m diameter) & Starship or equivalent (9 m diameter) & Starship or equivalent (9 m diameter)\\\hline
         Total Mass &  $\leq$ 25,000 kg & $\leq$ 37,500 kg & $\leq$ 37,500 kg\\\hline
         Minimum Aperture Inscribed Diameter&  6 m (7.2 m OD) & 6 m & 8 m\\\hline
         Telescope Configuration&  Off-axis& Off-axis & On-axis\\\hline
         Telescope Deployment Approach&  PM: JWST-like wings. SM: Deployable. &PM: Fixed. SM: Deployable& PM: JWST-like wings. SM: Deployable\\\hline
         Coronagraph CECs&  4-channel Optical and NIR, 1 mm 96 $\times$ 96 pitch DMs & 4-channel Optical and NIR, 1 mm 96 $\times$ 96 pitch DMs & 4-channel Optical and NIR, 1 mm 96 $\times$ 96 pitch DMs \\\hline
         Camera/Guider Description&  2 guider instruments at opposite ends of FOV.& 2 guider instruments at opposite ends of FOV & 2 guider instruments at opposite ends of FOV\\\hline
         Number of Instrument Bays&  3 + 2 guiders + 1 empty & 3 + 2 guiders + 1 empty  & 3 + 2 guiders + 1 empty \\\hline
 Mirror assumptions& ULE, hexagonal segments $<$ 1.8 m, operates at 20$^{\circ}$C. & ULE, keystone segments $\le$ 3 m, operates at 20$^{\circ}$C. & ULE, keystone segments $\le$ 1.8 m, operates at 20$^{\circ}$C.\\\hline
 Driving detector temperatures& 65K NIR APD (in coronagraph) with passive cooling&65K NIR APD (in coronagraph) with passive cooling & 65K NIR APD (in coronagraph) with passive cooling\\\hline
 Roll requirements& $\pm$22.5$^{\circ}$ & $\pm$22.5$^{\circ}$ &$\pm$22.5$^{\circ}$\\\hline
 Field of Regard& $\pm$45$^{\circ}$ pitch&$\pm$45$^{\circ}$ pitch &$\pm$45$^{\circ}$ pitch\\\hline
 Serviceability strategy& SIs, SC, and refueling&SIs, SC, and refueling &SIs, SC, and refueling\\ \hline
    \end{tabularx}
    \caption{EAC 1 - 3 Design Configuration Parameters. OD: Outer diameter. PM: Primary mirror. SM: Secondary Mirror. NIR: Near-infrared. DM: Deformable mirror. FOV: Field of View. ULE: Ultra-low expansion glass. APD: Avalanche photodiode. SI: Science Instrument. SC: Spacecraft. }
    \label{tab:eac123_tab}
\end{table}

The team did not choose to study a monolith for several reasons. First, monoliths of the size needed to accomplish many of the \ac{hwo} science cases lack flexibility for launch vehicle and aperture size (see Sec.~\ref{sec:science}).  Second, mirror stiffness scales non-linearly with diameter and a 6 - 8~m class monolith would be extremely low stiffness which both impacts gravity distortion (and the residual wavefront error) and stability (any instability like thermal changes through mounts would create large distortions). A low stiffness mirror would also create major new challenges for surviving the launch environment and would require a custom snubbing architecture that does not have flight heritage.  On top of this, \ac{fuv} coating uniformity is more challenging with size and would be extremely difficult to meet the requirements necessary for high-contrast imaging ($\lesssim$~3$\%$ variations, John Krist, priv. comm.). The risks associated with large monoliths are high during development due to the susceptibility of damage from mishandling. Finally, the facilitization does not exist for these large mirrors in the US using ultrastable materials.  Segmented systems have their own challenges and are sensitive to tip, tilt and piston stabilization which can be achieved with active metrology but require early investment.  In addition, diffraction effects and discontinuities from segmented mirrors slightly lower throughput but the expectation is that the segmented performance degradation is relatively low impact based on modeling, though this needs to be demonstrated on testbeds (\textcolor{red}{Belikov et al. 2026, this issue}).  Mirrors with exquisitely small edges have been made for ground telescopes (Benjamin Gallagher, priv. comm.), and the gaps between mirrors can be limited to just a handful of millimeters.  \ac{jwst} has demonstrated space-based wavefront sensing and control, hexapod controls, and integration and testing, making a segmented approach a mature method from a system perspective.

All of the \ac{eac} architectures were chosen have external starshade compatibility, but starshades are not baselined.  Since starshades are most suited for characterization and lack flight heritage, \ac{htmpo} is currently considering a starshade as a second generation capability that could be added, particularly for the \ac{nuv} where it does not need to be as large (a \ac{nuv} starshade would be closer to 30-35~m, more consistent with current starshade technology developments). 

\begin{table}
    \centering
    \begin{tabularx}{\textwidth}{X X}\hline
         \multicolumn{2}{|c|}{\textbf{Telescope}} \\\hline
         Diameter & $\sim$6 - 8 m (inner)\\
         Bandpass & $\sim$100 - 2500 nm \\\hline
         \multicolumn{2}{|c|}{\textbf{Coronagraph}}\\\hline
         Bandpass & $\sim$450 - 1700 nm\\
         Contrast & $\le$1 $\times$ 10$^{-10}$ \\
         R ($\lambda/\Delta\lambda$) & Vis: $\sim$140\\
                                     & NIR: $\sim$40\\
        Bandpass & 20\% (goal) \\\hline
        \multicolumn{2}{|c|}{\textbf{High-Resolution Imager}}\\\hline
        Bandpass & $\sim$200 - 2200 (TBD) nm \\
        Field of view & $\sim$3' $\times$ 2' \\
        60$+$ science filters and grisms\\
        High-precision astrometry\\\hline
        \multicolumn{2}{|c|}{\textbf{UV Multi-Object Spectrograph}}\\\hline
        Bandpass & $\sim$90 - 700 nm\\
        Field of view & $\sim$ 2' $\times$ 2'\\
        Apertures & $\sim$ 840 $\times$ 420\\
        R ($\lambda/\Delta\lambda$) & $\sim$500 - 60000\\\hline
        \multicolumn{2}{|c|}{\textbf{Other Science Instruments}}\\\hline
        \multicolumn{2}{l}{NUV coronagraph, NUV starshade, UV/VIS Integral Field Spectrograph, Spectropolarimeter}\\\\
    \end{tabularx}
    \caption{Instrument and telescope specification summary for \ac{eac}1-3; specifications are under development for \ac{eac}4 and \ac{eac}5.}
    \label{tab:placeholder}
\end{table}

Through the \ac{eac} design exercise, preliminary optical layouts of the telescope and instrument concepts were developed, which provided the necessary inputs for mechanical packaging. The mechanical team conducted multiple packaging iterations for each \ac{eac} to accommodate: (1) a large primary mirror and stiff secondary mirror tower, (2) four science instruments and two guiders, (3) a large telescope baffle for micrometeorite protection, and (4) spacecraft, radiator sunshades, and solar arrays, into at least two launch vehicle fairings. Lessons learned from \ac{jwst} informed the need for a large baffle that protects against both frequent, low energy ($\le$1 J) and rarer, higher energy micrometeorites \cite{Telfer2024}. Additional work on this topic can be found in \cite{Gialluca2025}. Further, with each successive \ac{eac} design, we not only developed a new architecture configuration but also refined and resolved issues identified in previous \ac{eac} designs. 

After completing the \ac{eac}-1 design, the modeling team generated physics-based models to enable end-to-end modeling and system-level performance evaluation, while the design team worked in parallel on additional architectures. Section 4 presents the top-level approach to system analysis, with additional details provided in \textcolor{red}{Liu et al., 2026, this issue}. 

\section{Integrated Modeling Approach and Results} 

\begin{figure}
\begin{center}
\begin{tabular}{c}
\includegraphics[height=8cm]{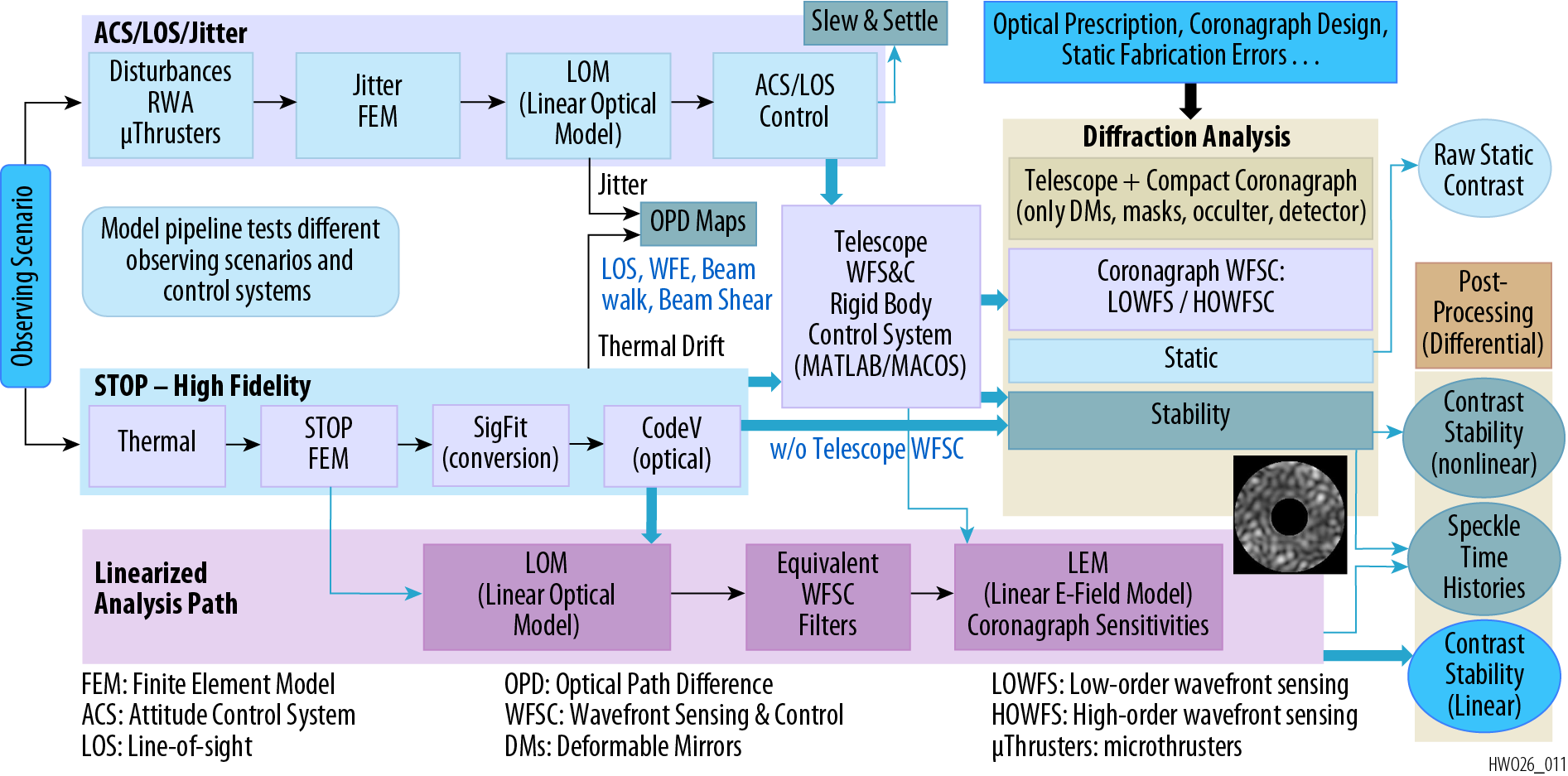}
\end{tabular}
\end{center}
\caption 
{ \label{fig:HWO-IMPipeline}
Block diagram showing the integrated modeling pipeline for the HWO performance analyses through the coronagraph. This pipeline structure builds upon significant integrated modeling efforts from \ac{jwst} and \ac{ngrst}. End-to-end integrated modeling simulations through this flow have been carried out.} 
\end{figure}

\ac{im} on \ac{hwo} builds upon the capabilities successfully demonstrated on JWST\cite{Levine2023} and Roman\cite{Liu2025, Krist2023}. \ac{im} is the analytical process by which multi-disciplinary engineering models of the Observatory system are exercised in a variety of sequences depending on the end-to-end performance to be estimated. This is what we refer to as the multi-thread modeling pipeline. Each discipline step in the thread requires a discipline-specific model which can be exercised using either commercial or custom developed analysis tools. All disciplines models are consistent with the mechanical geometry and the coordinate systems defined in the \ac{cad} model. Component and data interfaces between the models are strictly defined to enable automated data flow through the pipeline across different analysis teams and without the need for users in the loop. For \ac{hwo}, the \ac{im} pipeline for the coronagraph instrument is illustrated in Fig~\ref{fig:HWO-IMPipeline}. The main branches of the HWO pipeline are the \ac{acs}~/~\ac{los}~/~Jitter thread at high temporal frequencies and the thermal drift thread at low temporal frequencies, also commonly referred to as the \ac{stop} thread. Both threads can then be fed through the wavefront sensing control and diffraction threads to estimate nonlinear coronagraph electric fields, raw contrast and contrast stability which are compared to the top-level coronagraph requirements in the \ac{frn} error budget.

The jitter thread takes input models of \ac{acs} mechanism disturbances; for \ac{hwo}, we are investigating micro-thrusters and reaction wheels. For \ac{stop}, an \ac{os} defines the sequence of roll and pitch maneuvers of the observatory which affect the thermal environment and stability. The jitter and \ac{stop} threads produce optical figures of merit such as alignment, figure wavefront error, or beamwalk which are actively controlled through rigid body telescope segment correction, fast steering mirror alignment correction, and deformable mirror figure correction. Active thermal control of the primary mirror assembly and the coronagraph bench also contribute to improving the stability of the figures of merit and are included in the models. For the coronagraph, the predicted optical performance from jitter and/or \ac{stop} is then run through the diffraction analysis thread to generate the coronagraphic speckle image and \ac{psf} from which static raw contrast and contrast stability are estimated.

Because these analyses are nonlinear and computationally intensive, a linearized analysis path can be exercised for rapid evaluation of design trades. The \ac{im} pipeline for the other \ac{hwo} instruments, such as \ac{hri}, will use essentially the same observatory models and analysis threads as for the coronagraph but will propagate through each instrument’s individual optical path to evaluate instrument-specific metrics at the detectors such as \ac{ee}, \ac{ee}~Stability and Strehl Ratio. 

For the first \ac{im} cycle for \ac{hwo}, \ac{eac}1 provided the opportunity to pipeclean the process, which entails creating and verifying each of the discipline models and their respective data interfaces in the various threads of the pipeline. The \ac{eac}1 models, while very detailed for this phase of the mission, do not incorporate all features required for high-fidelity determination of performance. Simplifying or optimistic assumptions are incorporated into the \ac{eac}1 models, such as enforced temperatures at boundary nodes in lieu of active thermal control, ideal masks and deformable mirrors without fabrication errors or drift, and optimistic sensor noise in the metrology control systems. While the results appear to be within reach of the stability goals for exoplanet detection and characterization, no assessment is being made at this point against error budgets until the designs and models mature.

Prior to generating performance results, the models go through a rigorous verification process to ensure that the pipeline is computationally ready to generate end-to-end simulations, that there are no inconsistencies across the disciplines and interfaces, and that the models generate the expected results for a pre-determined set of analysis cases. When possible, model verification is performed by two independent teams to cross-check the results. Model validation, which is different than verification, measures the accuracy of a model to predict the right physics in the hardware. Model validation is a key activity of the technology program, whereby the fidelity of the models will improve commensurate with the progress of the \ac{trl} and delivery of flight-like hardware against which to build and test models. Just like \ac{jwst} and \ac{ngrst}, \ac{hwo} will define a model validation roadmap which will define each of the hardware configurations and tests used to improve the accuracy and fidelity of the models. A novel uncertainty quantification methodology is also being developed to assess the statistical confidence of the integrated model predictions to more robustly quantify margin of the system requirements, which does not solely rely on bounding worse case scenarios and individual discipline material property variations \cite{Levine2023}.

\begin{figure}
\begin{center}
\begin{tabular}{c}
\includegraphics[height=6.5cm]{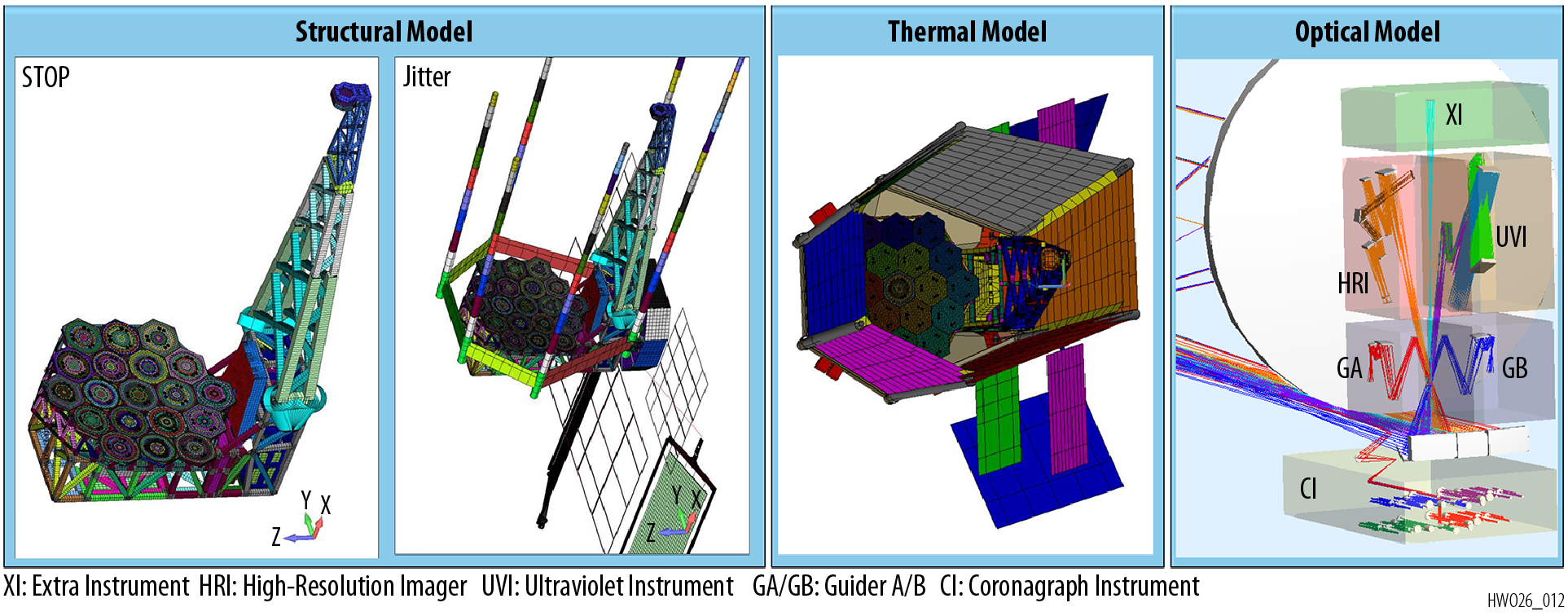}
\end{tabular}
\end{center}
\caption 
{ \label{fig:HWO-STOP}
Graphical representation of the various discipline models for \ac{eac}1 that are used by the HWO integrated modeling pipeline to predict the system-level performance and identify areas that require engineering refinement.} 
\end{figure}

Examples of the various \ac{eac}1 discipline models are shown in Figure~\ref{fig:HWO-STOP}. While each discipline model appears to be different, they are in fact all consistent with the baseline geometry defined in the \ac{cad} and the optical models. They are also consistent with the overall materials and properties definition. Only components which impact the engineering discipline performance are modeled in each step of the pipeline. For instance, in Figure~\ref{fig:HWO-STOP}, the thermal model represents the outer barrel assembly, the spacecraft, and the sunshade which drive the thermal stability of the observatory. However, the \ac{stop} model only uses temperatures along the telescopes and instrument payloads to determine the stability of the alignment and wavefronts.

As was the case for \ac{jwst} and \ac{ngrst}, \ac{hwo} will rely on \ac{im} for final verification prior to launch and during commissioning, since end-to-end system testing to the required picometer stability will be challenging on the ground. Hence this \ac{im} process will endure for the duration of the project lifecycle with improved, higher fidelity models as the models are validated against test data and designs are refined. It is important to note that the success of the \ac{im} pipeline also resides in the carefully orchestrated collaboration of the individual discipline analysts. On \ac{jwst} and \ac{ngrst}, it was not uncommon for the key analysts to support the mission through its lifecycle across decades. \ac{hwo} is fortunate to benefit from the expertise of these \ac{jwst} and \ac{ngrst} analysts to set the foundations of the \ac{im} process and train the next generation.

Once the \ac{im} pipeline process is successfully demonstrated, it was applied to investigate possible design options and thermo-optical control strategies for \ac{eac}1. The results shown here are for preliminary engineering demonstrations of the modeling pipeline only with idealized assumptions. A snapshot of the performance results are presented here and a more detailed descriptions are provided in \textcolor{red}{Liu et al., 2026, this issue}.

\begin{figure}
\begin{center}
\begin{tabular}{c}
\includegraphics[height=5.75cm]{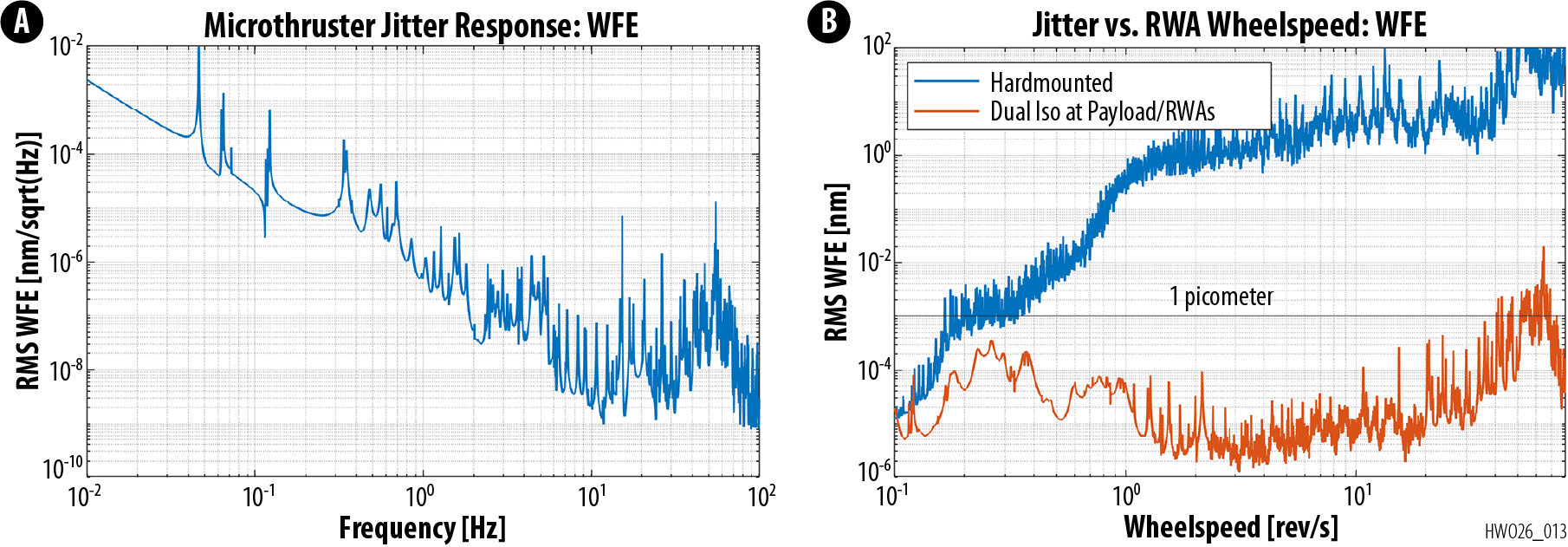}
\end{tabular}
\end{center}
\caption 
{ \label{fig:HWO-jitter}
Comparison of the wavefront jitter performance between attitude control using micro-thrusters (left) or traditional reaction wheel assemblies, either hard-mounted or on an isolation platform (right).} 
\end{figure}

Figure~\ref{fig:HWO-jitter} illustrates the output of the jitter pipeline which compares the figure \ac{wfe} resulting from attitude control using either a set of micro-thruster disturbances or eight reaction wheels mounted on a pallet with or without isolation. Such analyses provide valuable insight to assessing design trades and impacts to performance objectives.

\begin{figure}
\begin{center}
\begin{tabular}{c}
\includegraphics[height=20cm]{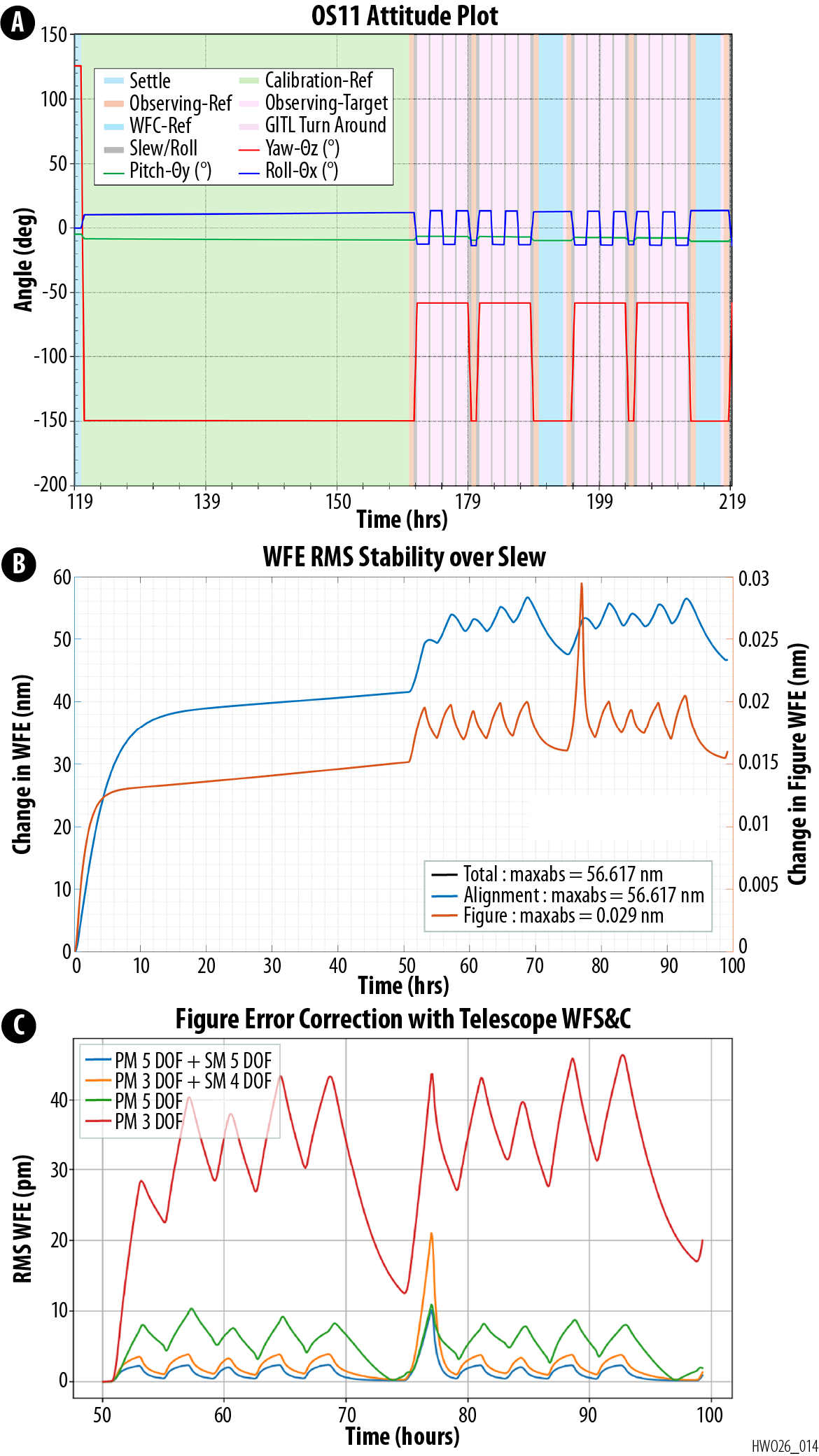}
\end{tabular}
\end{center}
\caption 
{ \label{fig:CoronagraphOS}
Representative observing scenario simulation on the wavefront error transient time history with and without telescope figure control. WFC: Wavefront control. GITL: Ground-in-the-loop. WFE: Wavefront error. WFS\&C: Wavefront sensing and control. PM: Primary Mirror, SM: Secondary Mirror.  DOF: Degrees of freedom.} 
\end{figure}

Similarly, Figure~\ref{fig:CoronagraphOS} illustrates representative outputs from simulating the orbital maneuvers for a representative coronagraph \ac{os} to detect and image an exoplanet. The \ac{os} drives the thermal stability of the observatory as a series of roll and pitch maneuvers between a nearby star used for wavefront control followed by angular differential imaging on a target star. Wavefront stability at the picometer-level is needed to detect the planet from the background noise speckles. A representative maneuver sequence for \ac{os}~1 is shown in Figure~\ref{fig:CoronagraphOS} along with the predicted figure wavefront stability with and without active rigid body control of the segments.

\begin{figure}
\begin{center}
\begin{tabular}{c}
\includegraphics[height=5cm]{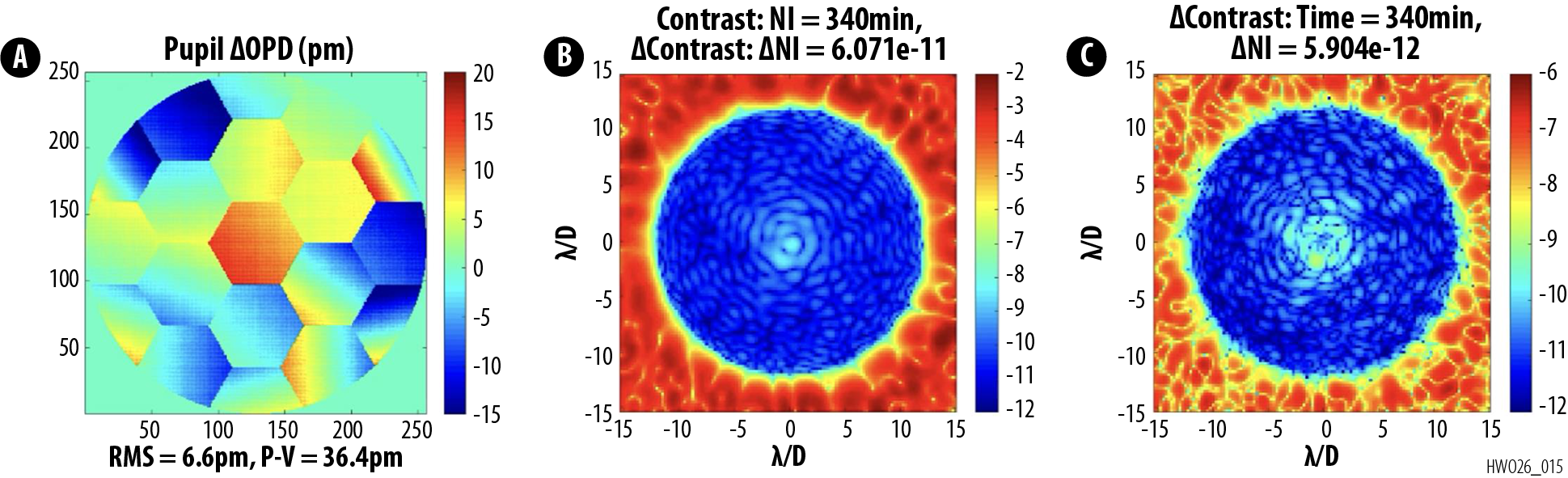}
\end{tabular}
\end{center}
\caption 
{ \label{fig:ContrastStability}
Sample timestep of the observing scenario contrast and contrast stability time histories with rigid-body metrology and deformable mirror control. The integrated modeling timeseries evolution of these scenes can be used to evaluate post-processing gains.} 
\end{figure}

In the last step of the pipeline, the optical performance from the \ac{stop} thread is processed through the telescope segment rigid-body control and coronagraph deformable mirror control to predict time histories of the contrast and contrast stability as the observatory progresses through the \ac{os}~1. A snapshot in time of the electric field diffraction speckles and stability is illustrated in Figure~\ref{fig:ContrastStability}. Eventually these speckle electric fields can be further post-processed with methods such as \ac{adi} to further improve the contrast and planet detection from the speckles. 

\section{Architectural Refinements Based on Trades} 

After successfully completing designs for \ac{eac}s 1-3 and integrated modeling for \ac{eac}1, we identified key design challenges and documented lessons learned\cite{Feinberg2024}. From this experience, we developed configuration parameters for the next set of \ac{eac}s. These parameters serve as top-level architecture definitions and are shown in Table~\ref{tab:EAC4-5}.

For \ac{eac}s 4-5, design goals include dual rocket compatibility, which establishes mass and volume constraints. The aperture size is notably increased from the first three \ac{eac}s to provide higher throughput and exoplanet yield. Four instruments are planned, with a placeholder instrument that can be added post-launch through servicing. There was strong scientific interest in high spatial resolution spectroscopy in the UV to study galaxy growth and the evolution of the elements, which led to the addition of a UV integral field spectrograph (see \S~\ref{subsec:GalaxyGrowth} and \S~\ref{subsec:EvolutionElements}). Instrument design goals will be established by the science and engineering teams. The \ac{eac}4/5 design can observe targets within a \ac{for} of $\pm$45$^{\circ}$ pitch (toward and away from the Sun) and $\pm$22.5$^{\circ}$ roll about the observatory optical axis. Servicing assumptions include instrument changeouts, spacecraft bus or component replacement, and refueling.

\begin{table}
    \centering
    \begin{tabularx}{\textwidth}{X X X}\hline
         \textbf{Configuration Parameter}&  \textbf{EAC4}& \textbf{EAC5}\\\hline
         Launch Vehicle Compatibility&  Starship, Next Generation Blue Origin, or SLS (dual compatibility)& Starship, Next Generation Blue Origin, or SLS (dual compatibility). Space integration TBR.\\\hline
         Total Mass &  $\leq$ 25,000 kg & $\leq$ 37,500 kg\\\hline
         Minimum Aperture Inscribed Diameter&  6.5 - 7 m& 8 - 8.5 m\\\hline
         Telescope Configuration&  Off-axis& Off-axis\\\hline
         Telescope Deployment Approach&  PM: Fixed. SM: Deployable& PM and SM: Deployable\\\hline
         Coronagraph CECs&  2 parallel channels, 2 Visible or Visible + NIR.
2 64 $\times$ 64 DMs with 1 mm pitch& 3 parallel channels, 2 Visible and 1 NIR. 
4 DMs: 96 $\times$ 96 with 1~mm pitch for VIS, 64 $\times$ 64 with 1~mm pitch for NIR1, and 64 $\times$ 64 with 0.4~mm pitch for NIR2. \\\hline
         Integral Field Unit&  FUV/NUV& FUV/NUV\\\hline
         Camera/Guider Description&  4 quadrant design with NIR. Includes guiding.& 4 quadrant design with NIR. Includes guiding.\\\hline
         Number of Instrument Bays&  4+1 Empty& 4+1 Empty\\\hline
 Mirror assumptions& ULE, hexagonal segments $<$ 1.8 m, operates at 20$^{\circ}$C.&ULE, hexagonal segments $<$ 1.8 m, operates at 20$^{\circ}$C.\\\hline
 Driving detector temperatures& 65K NIR APD (in coronagraph) with passive cooling&65K NIR APD (in coronagraph) with passive cooling\\\hline
 Roll requirements& $\pm$22.5$^{\circ}$
&$\pm$22.5$^{\circ}$\\\hline
 Field of Regard& $\pm$45$^{\circ}$ pitch&$\pm$45$^{\circ}$ pitch\\\hline
 Serviceability strategy& SIs, SC, and refueling&SIs, SC, and refueling\\ \hline
    \end{tabularx}
    \caption{EAC 4/5 Design Configuration Parameters. OD: Outer diameter. PM: Primary mirror. SM: Secondary Mirror. NIR: Near-infrared. DM: Deformable mirror. FUV: Far ultraviolet. NUV: near ultraviolet. ULE: Ultra-low expansion glass. APD: Avalanche photodiode. SI: Science Instrument. SC: Spacecraft. }
    \label{tab:EAC4-5}
\end{table}

\begin{figure}
\begin{center}
\begin{tabular}{c}
\includegraphics[height=7cm]{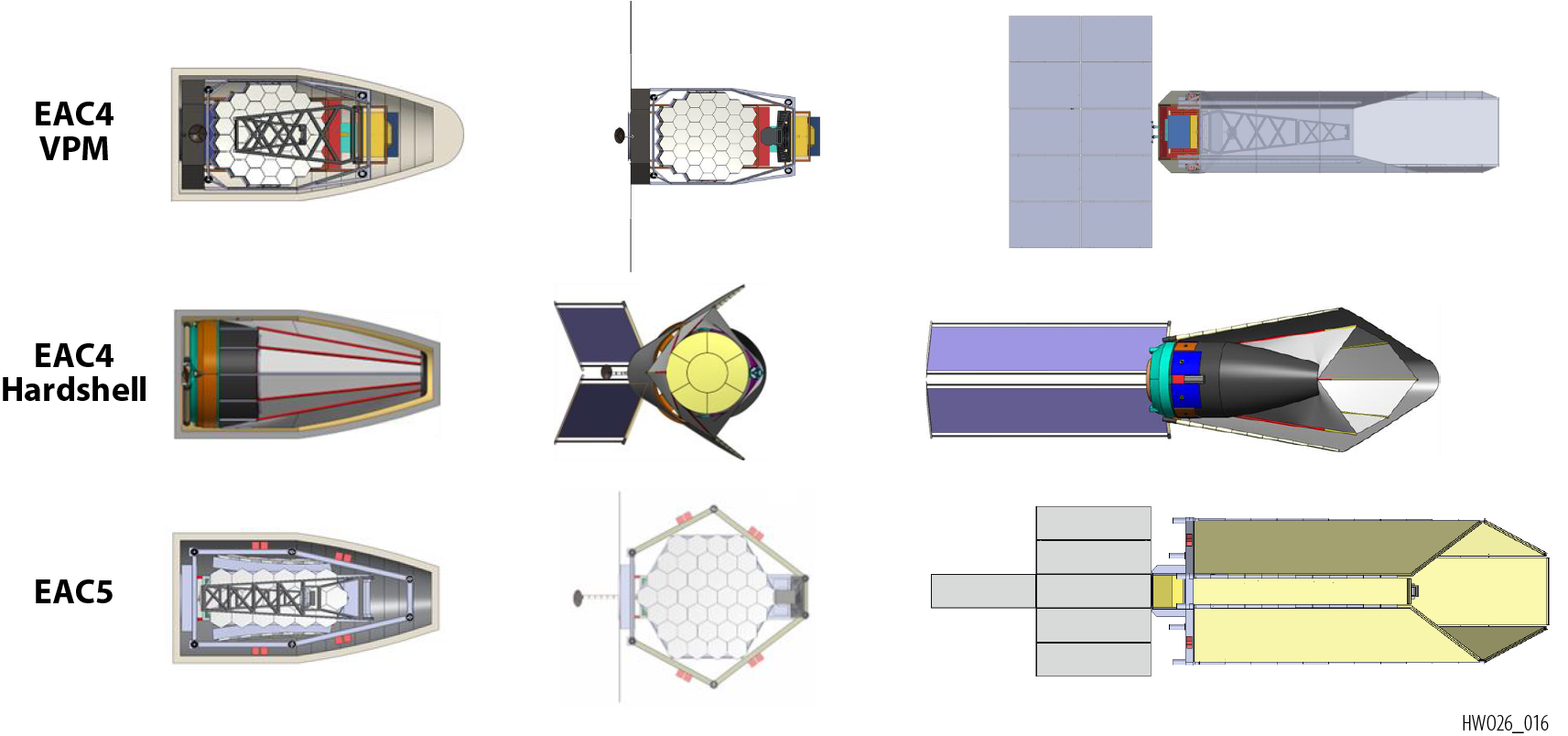}
\end{tabular}
\end{center}
\caption 
{ \label{fig:EACs45}
Two EAC4 and one EAC5 concept. Similar to Figure~\ref{fig:EACs123}, the left column shows the stowed launch configuration for each architecture and its fit within the representative Starship fairing. The middle column is looking down the telescope boresight, and the right column shows the top-down view.} 
\end{figure}

Based on configuration parameter definitions, we have developed several initial design concepts. As shown in Figure~\ref{fig:EACs45} top row, the \ac{eac}4 ``Hardshell'' concept reduces telescope barrel deployment complexity. Figure~\ref{fig:EACs45} middle row presents an alternative \ac{eac}4 design featuring a stowed \ac{vpm} in the launch vehicle. Figure~\ref{fig:EACs45} bottom row displays the \ac{eac}5 initial design concept with an 8-m inscribed diameter primary mirror. The \ac{eac}4 designs prioritize simplified deployments while maximizing primary mirror size, whereas \ac{eac}5 focuses on accommodating a larger primary mirror with a minimum 8-m inscribed diameter.

A key distinction between approaches is that \ac{eac}4 concepts avoid folding the primary mirror, while the \ac{eac}5 configuration requires folded primary mirror sides at launch, similar to \ac{jwst}'s packaging design. Further, to accommodate the larger primary mirror in \ac{eac}5, the telescope barrel deployment mechanism requires at least two degrees of freedom; this allows deployment posts to extend outside the primary mirror before deploying along the telescope axis. Following evaluation of these initial designs, we will select two configurations, designated as the final \ac{eac}4 and \ac{eac}5, for further detailed study.

\begin{figure}
\begin{center}
\begin{tabular}{c}
\includegraphics[height=7cm]{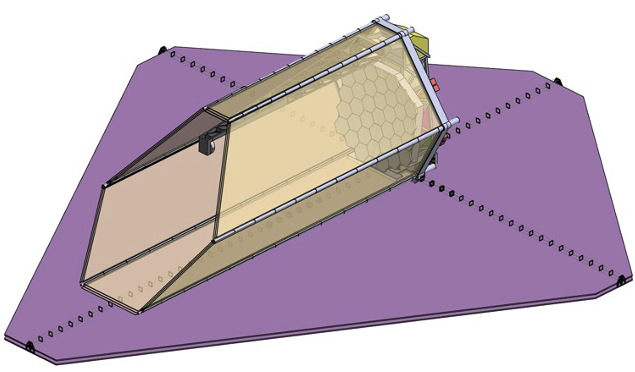}
\end{tabular}
\end{center}
\caption 
{ \label{fig:EAC5_flatshade}
EAC5 concept shown with a notional flat sunshield. The flat sunshield improves the thermal performance and stability as the observatory points within its field of regard.} 
\end{figure}

In addition to designing two new architectures, several key engineering trades will be performed in parallel across the observatory and instruments. Key observatory-level trades include the evaluation of using a flat sunshield as a means to improve the observatory thermal stability (see Figure~\ref{fig:EAC5_flatshade}) and whether in-space integration of the telescope barrel and/or sunshields could simplify the overall observatory design, assessing the required level of servicing autonomy. Telescope performance trades will compare various \ac{wfsc} and actuation methodologies, assessed while tracking residual closed-loop errors using \ac{im}, and evaluate vibration and isolation options to mitigate jitter (micro-vibration). A study of detector cooling options, including cryogens and cryocoolers, will be carried out in addition to the passive cooling assumed in the \ac{eac}s to date. Finally, several coronagraph observing scenarios and post processing methodologies will be evaluated and modeled with the goal of relaxing design complexity and stability requirements.

With additional design and modeling experience from \ac{eac}4/5, we will be ready to select a point design for \ac{eac}6. The team will focus on maturing this design to demonstrate feasibility and complete requirement derivation in preparation for \ac{mcr}.

\section{Technology Maturation Approach}
\ac{hwo}'s technology development needs are organized into three general categories, or ``tracks'': Coronagraph System Technologies, Ultra-stable Telescope System Technologies, and High-sensitivity Ultraviolet and Visible Instrument Technologies. Within each track, individual technology lanes describe the critical technology needs to enable the technology system. In some cases, multiple candidate technologies of different maturity may address a particular lane, offering the opportunity for parallel development and down-selecting to the best performing candidate. Table~\ref{tab:tech_lanes} shows the current set of technology lanes for each track.

\begin{table}
    \centering
    \begin{tabularx}{\textwidth}{|X|X|X|}\hline
         \textbf{Track 1:} \textit{Coronagraph System Technologies}&  \textbf{Track 2:} \textit{Ultra-stable Telescope Technologies}& \textbf{Track 3:} \textit{High-sensitivity Ultraviolet and Visible Instrument Technologies}\\\hline
         Starlight Suppression&  Ultra-stable Mirrors& Far-UV Mirror Coatings\\\hline
         Contrast Stabilization&  Ultra-stable Structures& Near-UV/Visible Detectors\\\hline
         Deformable Mirrors&  Thermal Control Systems& Far-UV Detectors\\\hline
         Low-noise/Noiseless Detectors (visible and NIR)&  Telescope Wavefront Sensing $\&$ Control& Multi-object Selection/ Integral Field Units\\\hline
         Spectroscopy&  Low-Disturbance Systems& UV Gratings $\&$ Filters\\\hline
         Post-processing&  Deployable Systems& \\\hline
         Near-UV Capability&  & \\ \hline
    \end{tabularx}
    \caption{The \ac{hwo} technology program is organized into tracks (columns) within lanes (rows) for each technology.}
    \label{tab:tech_lanes}
\end{table}

Despite this organization, the technologies within each track are not isolated from those in other tracks, and many cross-track interfaces exist. For example, coronagraph system and ultra-stable telescope system technologies are closely related in that coronagraph performance may require a specific telescope stability, and the telescope stability that is actually feasible may levy additional requirements on specific coronagraph technologies. Similarly, it is known that some mirror coatings that enable far-ultraviolet reflectivity may create polarization or amplitude aberrations that affect coronagraph performance. Cooling requirements for some new detector technologies may have an impact on telescope stability. Managing these interfaces will continue to be a primary responsibility of the \ac{htmpo}.

To assess the readiness of candidate technologies for \ac{hwo}, we use the standard definition of \ac{trl} from NASA Procedural Requirement 7123.1 \cite{NPR7123}, augmented with additional details from NASA's Technology Readiness Assessment Best Practices Guide \cite{TRLBPGuide}. These documents provide clear guidance on identifying technology gaps versus engineering gaps and assessing the readiness of a technology for infusion into a flight mission. We further prioritize both engineering and technology gaps by assessing their importance and urgency. Importance gauges how enabling the technology is for \ac{hwo}:

\begin{itemize}
    \item \textbf{Threshold} technologies are those that are operationally critical to the threshold \ac{hwo} mission. If the threshold technologies are not matured, the \ac{hwo} mission cannot proceed.
    \item \textbf{Baseline} technologies are those that enable the baseline \ac{hwo} mission science objectives. Baseline technologies generally have offramps, such that if their development poses unacceptable cost or schedule risk, a state-of-the art backup could be used that would still enable at least the \ac{hwo} threshold science objectives. In such a scenario, the baseline technology would become a candidate for inclusion on a future servicing mission.
    \item \textbf{Enhancing} technologies are those that are not fundamentally necessary to enable the \ac{hwo} mission. Instead, enhancing technologies are those that would either substantially improve the science yield of the mission beyond the baseline objectives, or ease the development, implementation, test, and verification of the mission system.
\end{itemize}

Urgency gauges the impact of the technology’s development to either the pre-formulation critical path (i.e., readiness of the technology by \ac{mcr}), or to the mission design process:

\begin{itemize}
    \item \textbf{Critical} technologies are those that require substantially complex development or consist of long duration activities that must begin early. Critical technologies are also those where the performance must be well understood early enough to inform the mission design activities being conducted during pre-formulation.
    \item \textbf{Urgent} technologies are those that risk not achieving \ac{trl}~5 by \ac{mcr} or shortly thereafter. The specific performance capabilities of these technologies are not generally needed to inform pre-formulation mission design activities.
    \item \textbf{Long-term} technologies are those that can be matured later in the pre-formulation or formulation process as they pose low cost or schedule risk to the project.
\end{itemize}

\begin{figure}
\begin{center}
\begin{tabular}{c}
\includegraphics[height=10cm]{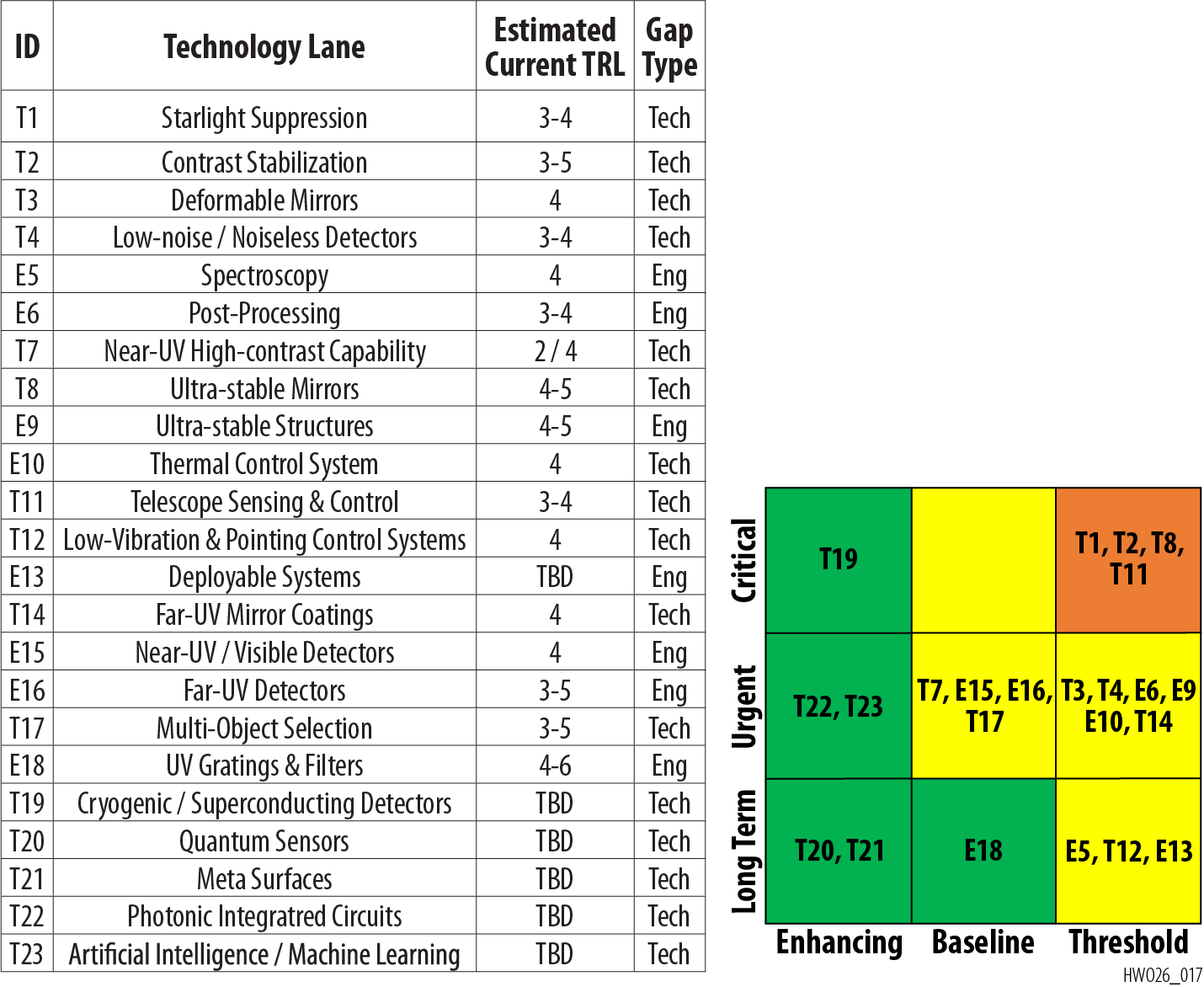}
\end{tabular}
\end{center}
\caption 
{ \label{fig:tech3x3}
Summary of all HWO technologies, estimates of their current TRLs, and their prioritization. The IDs in the table on the left are presented in a prioritized matrix on the right.} 
\end{figure}

Each technology's priority is reflected on a 3$\times$3 matrix. Figure~\ref{fig:tech3x3} shows the prioritization of all of HWO's gaps, including their categorization as either an engineering or technology gap, including potentially emerging, enhancing technologies.

\begin{figure}
\begin{center}
\begin{tabular}{c}
\includegraphics[height=9cm]{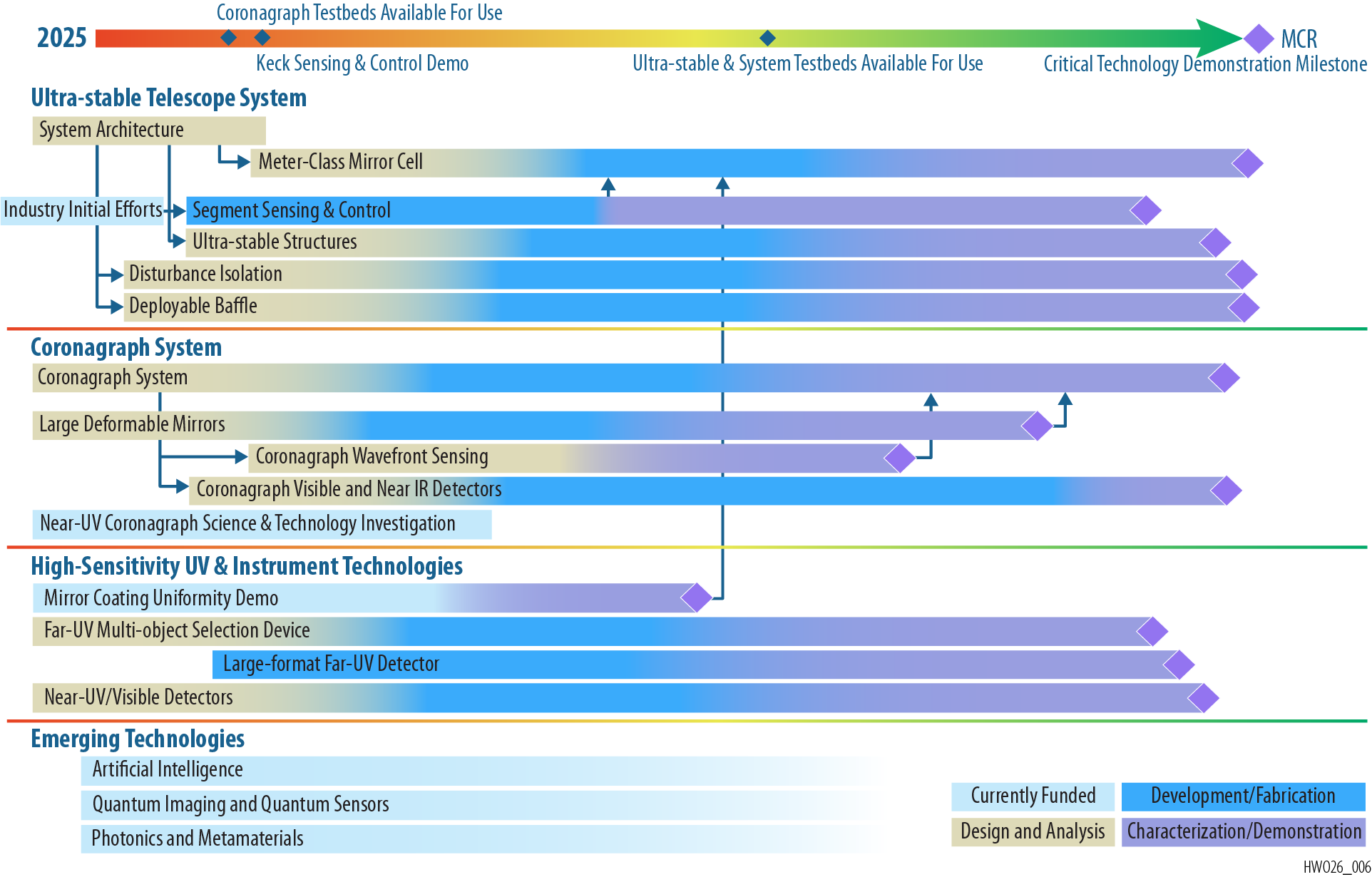}
\end{tabular}
\end{center}
\caption 
{\label{fig:TechnologyRoadmap}
The \ac{hwo} Technology high-level roadmap to \ac{mcr}. The technologies are broken down into their lanes, and the shading represents how technologies achieve TRL5 by \ac{mcr}. As technologies are matured, they can be incorporated into higher-level system demonstrations.} 
\end{figure}

A detailed technology development plan to mature each of the Baseline and Threshold technologies to \ac{trl}~5 prior to \ac{mcr} and evaluating enhancing technologies for potential inclusion in the mission has been completed. Figure~\ref{fig:TechnologyRoadmap} shows a high-level roadmap for the development of most of these technologies. Strategic investments with industry partners, government testbed facilities, academia, and small business partners will be used to execute this roadmap.

To manage technology investments, the team is using a risk-based approach described in the \ac{hwo} Technology Plan \footnote{Note the publicly-available \ac{hwo} Technology Plan has redacted all sensitive information.}. Each technology is placed on a technology matrix that factors in urgency and criticality.  The full \ac{hwo} Technology Plan, over 100 pages in length, was delivered to NASA Headquarters in April 2025 and has been used to guide the project investments.

\section{Testbed Demonstrations for Technology Maturation: Progress and Plans} 
Testbeds are critical for technology development and maturation, as they provide the means to ensure that critical components, subsystems, and systems demonstrate the required \ac{trl}. Two sets of testbeds are under development at \ac{gsfc} and \ac{jpl} that provide the community the environment and means to complete technology development for ultrastability and coronagraph development. 

\subsection{Ultra-Stable Structures Laboratory}
The \ac{ussl} has a history of making picometer accuracy dynamic tests over the spatial extent of test articles\cite{Saif2017, Saif2019}, and has been actively working towards achieving stability measurements at the required levels over the past few years. 

The \ac{ussl} is located at NASA's \ac{gsfc} in Greenbelt, Maryland. The room is acoustically isolated with sand-filled walls and acoustic tiles, and the heating, ventilation, and air conditioning system maintains the room at $\pm$0.5~K. There are two optical benches, including a 15.5' long $\times$ 5' wide optical bench with an invar top and bottom plate and honeycomb center; that optical bench has six active piezo actuators that isolate it from the ground at certain frequencies. The lab also has an 8’ $\times$ 4’ granite optical bench weighing $\sim$5070 pounds with no isolation underneath it. Both tables can accommodate modular optical arrangements that consist of the target of interest, an interferometer, and the supporting opto-mechanical setup. Two interferometers are used when taking measurements: a modified 4D~Technology Phasecam 6100 and a 4D~Technology HP4020\cite{Sitarski2026}.

\subsubsection{Early Laboratory Results: Towards Picometer Stability} \label{subsec:TowardsPicometerStability}
\begin{figure}
\begin{center}
\begin{tabular}{c}
\includegraphics[height=7cm]{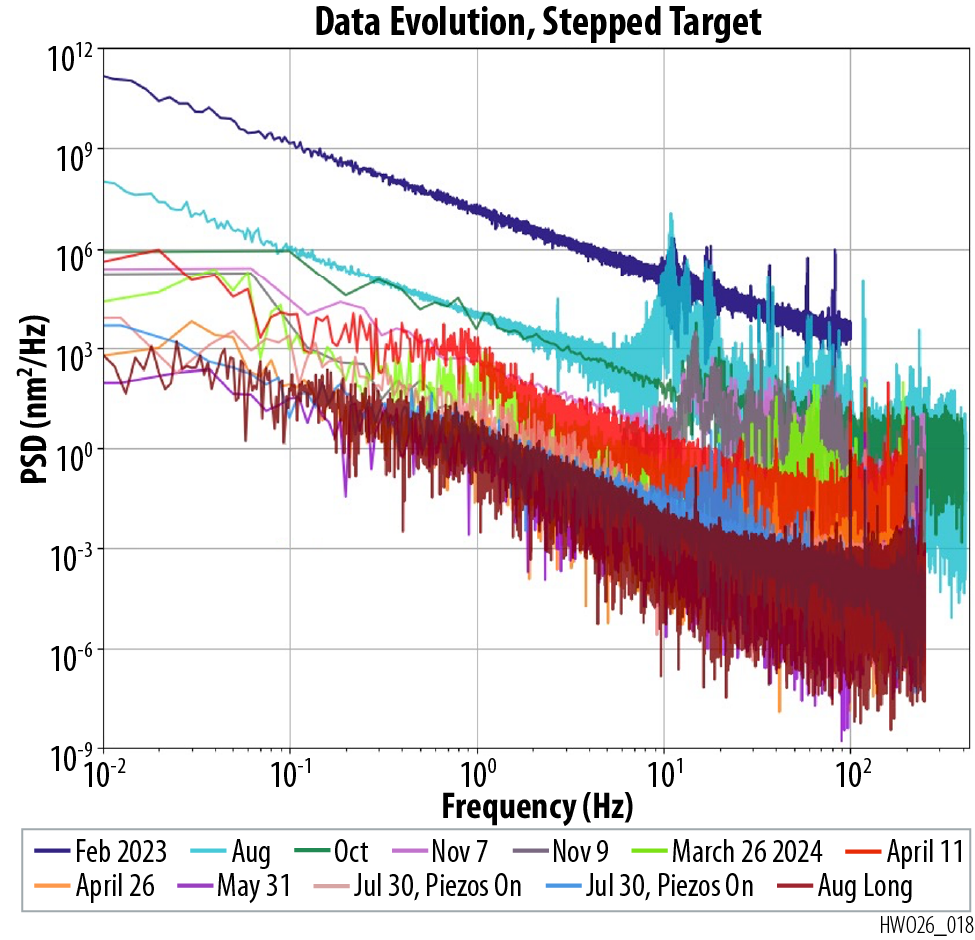}
\end{tabular}
\end{center}
\caption 
{ \label{fig:PSDev}
Power spectral density evolution for each of the optical setups the \ac{ussl} team tested between February 2023 and August 2024. The noise has decreased by 8 orders of magnitude in power, mainly due to rigidizing the mechanical connections between the various pieces of the setup and understanding underlying metrology errors.} 
\end{figure}

Between February 2023 and August 2024, the team was able to improve the overall noise in the system by eight orders of magnitude by mechanically rigidizing the system and further understanding various sources of noise. The power spectral density plot of global piston, the Zernike term that the team uses to assess overall system noise, is shown in Figure~\ref{fig:PSDev}. Low temporal frequency drift remain the largest source of noise.

\begin{figure}
\begin{center}
\begin{tabular}{c}
\includegraphics[height=10cm]{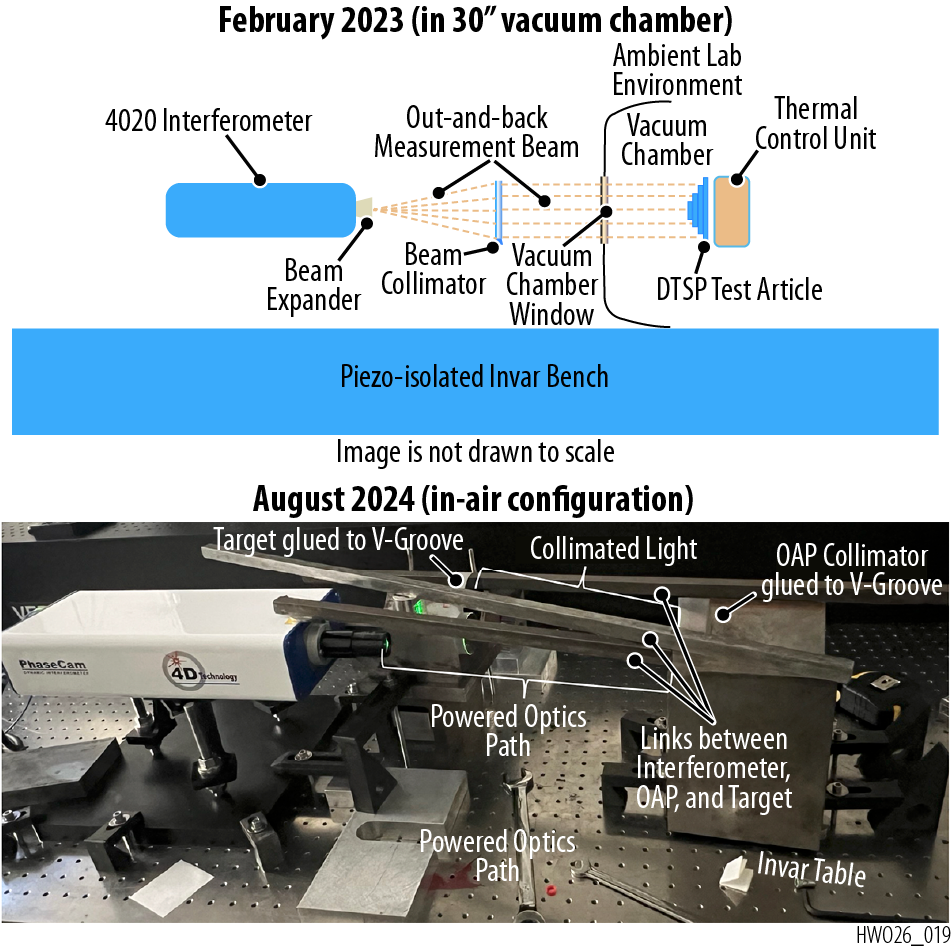}
\end{tabular}
\end{center}
\caption 
{ \label{fig:Configs}
Evolution of the optical setup in the \ac{ussl}, showing the February 2023 (top, dark blue line in Figure~\ref{fig:PSDev}) and the August 2024 setup (bottom, dark red line in Figure~\ref{fig:PSDev}). The primary changes were features to enhance mechanical rigidity.} 
\end{figure}

The largest impact to improving stability was enhancing the overall mechanical rigidization of the system. Figure~\ref{fig:Configs} shows the evolution of the setup from February 2023 to August 2024. While the early tests were performed in a 30" vacuum chamber with $\pm$75 mK/hour thermal stability\cite{Feinberg2022}, the lack of mechanical coupling between all parts of the optical system impacted stability measurements. The best results shown in Figure~\ref{fig:PSDev} are from a setup that is in-air but is mechanically rigid. The in-air setup resulted in low-spatial frequency relative variations on the surface of a 4” target that were $\geq$15 pm (weighted standard deviation) over all temporal frequencies when incorporating five, two-minute data sets. Low-temporal frequency drift within each 2-min interval limited the achievable precision. All the setups discussed in this section were located on the Invar bench.

\subsubsection{The Picochamber and Recent Metrology Additions}

\begin{figure}
\begin{center}
\begin{tabular}{c}
\includegraphics[height=7cm]{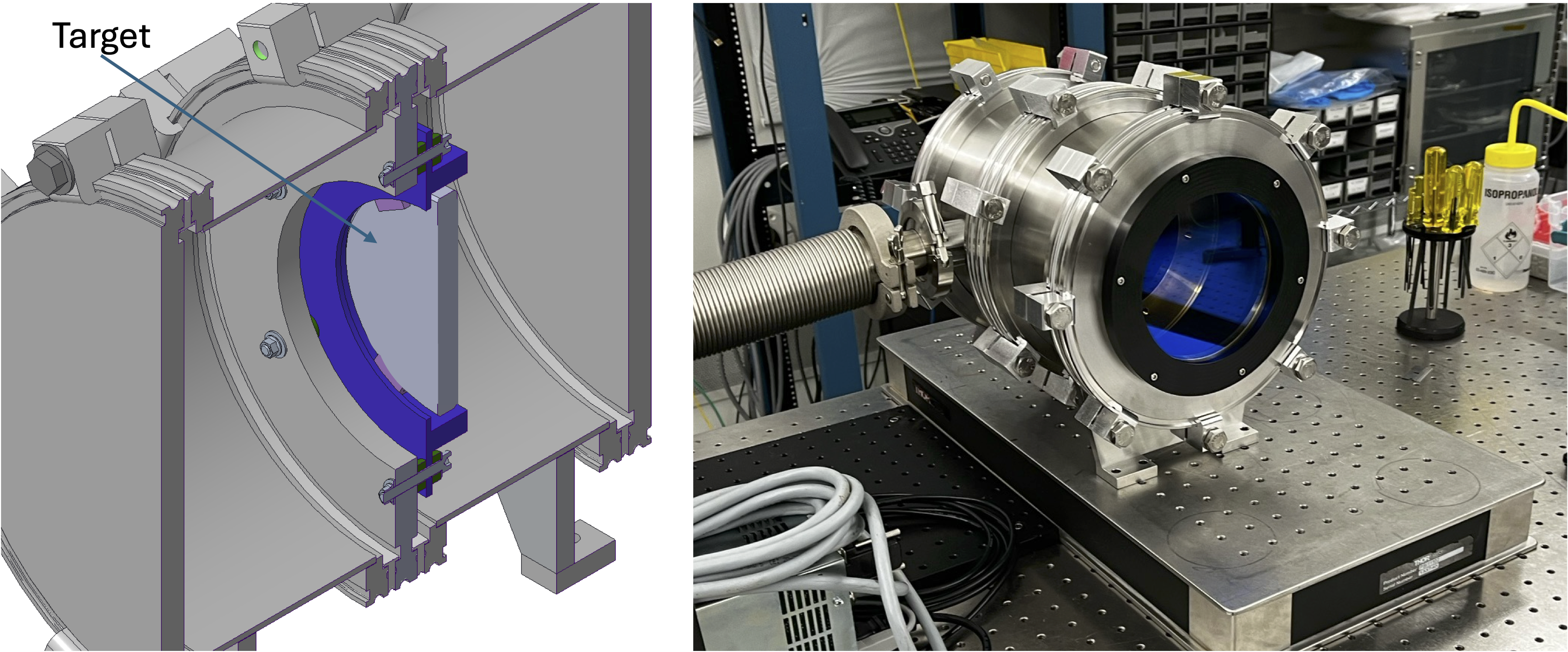}
\end{tabular}
\end{center}
\caption 
{ \label{fig:picochamber}
\textit{Left}: A cut-through of the picochamber shows the target mounted in a bezel that is mechanically coupled to the rest of the chamber, providing high rigidity. \textit{Right}: Picture of the picochamber in the \ac{ussl}.} 
\end{figure}

Building off stability lessons learned in the \ac{ussl} described in Section~\ref{subsec:TowardsPicometerStability}, the lab recently commissioned a small, 5" inner diameter vacuum chamber online that is designed to be rigid and stable. The chamber has a first natural frequency $>$100~Hz and is unique in that the target sits in a circular bezel that is incorporated into the center of the chamber, and the chamber itself is mechanically coupled to the measurement platform (see Figure~\ref{fig:picochamber}). This rigidizes the entire measurement system and ensures that the target is not independently vibrating within the vacuum chamber, which was likely the case in the 30” chamber (see Section~\ref{subsec:TowardsPicometerStability}). 

The picochamber can achieve $\pm$2 mK thermal control and be drawn down to micro-Torr level vacuum. The bezel is designed to accommodate 4 – 5” optics. The chamber is hot-biased under active thermal control with heaters outfitted on the outside of the chamber to stabilize the temperature inside the chamber\cite{Sitarski2025}.  

There is a clear impact of low temporal frequency ($<$0.1~Hz) drift on  measurement errors. These drifts cannot be isolated by commercially available isolators, but they can be sensed optically with autocollimators and with seismometers. Other environmental telemetry, such as pressure and temperature changes along the optical path, are also acquired to potentially correct for optical path difference errors\cite{Sitarski2022}.

\begin{figure}
\begin{center}
\begin{tabular}{c}
\includegraphics[height=8cm]{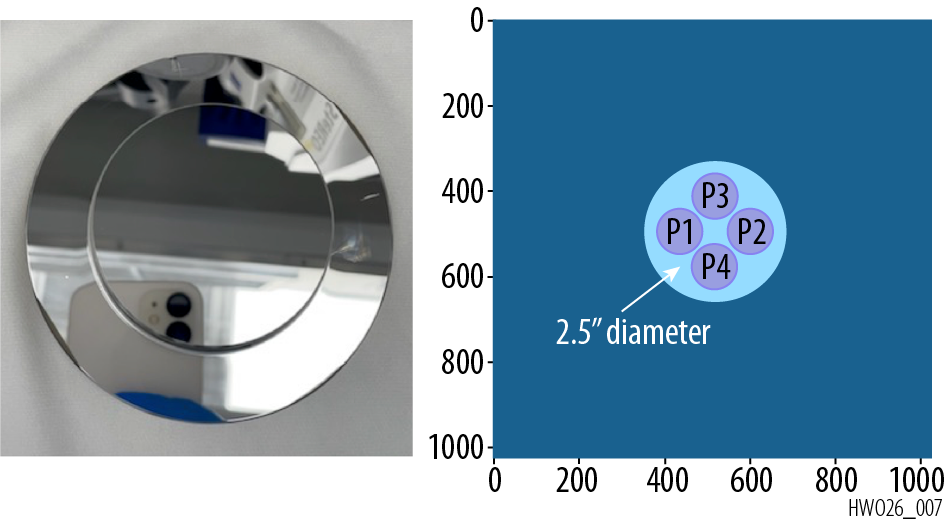}
\end{tabular}
\end{center}
\caption 
{ \label{fig:picochamberresults}
\textit{Left:} Image of the \ac{ule} test article, which has two reference surfaces for absolute calibration. \textit{Right:} Mask of the four subregions on the top surface of the \ac{ule} optic. The teal circle indicates the 2.5" region of interest.} 
\end{figure}

The team recently took data on a 4” \ac{ule} glass target. Twenty, 10-minute data sets were acquired on the same night. Twenty-eight global Zernike terms were subtracted from the inner 2.5” diameter of the target to subtract out global modes before dividing the same region into four patches. Twenty-eight Zernikes were in turn fit to those patches to determine the variation between each patch. The mask and resulting variations are shown in Figure~\ref{fig:picochamberresults} and Table~\ref{tab:picochamber_results}. When incorporating twenty data sets, relative differences were sensed at the single digit picometer level for piston, tip, and tilt with the picochamber. The target itself is a stepped \ac{ule} target (see left side of Figure~\ref{fig:picochamberresults}), so future work will include measuring the height differences at different temperatures to characterize the coefficient of thermal expansion for this particular grade of \ac{ule}. That measurement, combined with the relative difference measurements reported in Table~\ref{tab:picochamber_results}, demonstrate two types of measurements (relative and absolute) that must be demonstrated by various subsystems and systems to advance \ac{trl}. Additional results from other tests will be detailed in a forthcoming paper\cite{Sitarski2026}.

\begin{table}
    \centering
    \begin{tabular}{|c|c|c|c|}\hline
         &  \multicolumn{3}{|c|}{\textbf{Weighted $\sigma$ (picometers)}}\\\hline
         \textbf{0.01 - 10 Hz}&  \textbf{Piston}&  \textbf{Tip}& \textbf{Tilt}\\\hline
         P1 - P2&  2.07&  3.99& 8.81\\\hline
         P1 - P3&  2.62&  1.60& 3.92\\\hline
         P1 - P4&  2.85&  6.76& 5.09\\\hline
         P2 - P3&  4.61&  4.16& 4.40\\\hline
         P2 - P4&  6.38&  4.23& 8.08\\\hline
         P3 - P4&  4.63&  4.96& 5.41\\ \hline
    \end{tabular}
    \caption{Weighted standard deviation of relative differences on the same face of the optic when incorporating 20, 10-minute data sets.}
    \label{tab:picochamber_results}
\end{table}

\subsection{Mini-Metrology and Ultra-Stable Testbed}

The picochamber and other testing done in the \ac{ussl} is critical for understanding the sources of error that would impact picometer-level stability. However, those tests are small-scale (4 – 5”), and significant TRL advancement must be demonstrated on meter-class optics that are traceable to HWO-sized systems. The Mini-Metrology and Ultra-Stable Testbed (Mini-MUST) is a facility at GSFC that is designed to test up to 1-m class optics and systems to advance their TRLs to 5. It will be housed in the \ac{ciaf} at GSFC. The CIAF is a class 10,000 cleanroom that maintains thermal control within 0.5~K. Mini-MUST is situated on its on concrete pad. 

The Mini-MUST design relied heavily on integrated modeling and lessons learned in the USSL and other facilities to ensure that the system would be stable. The interferometer will not be able to operate within the vacuum chamber itself, so it sits outside but as close to the window as possible to minimize the air path. The interferometer and supporting optics sit on a rotopod that enables swift alignment. Within the vacuum chamber, two nested thermal shrouds ensure that $\pm$1~mK thermal control – necessary for picometer-level metrology – is achievable with 8 different thermal control zones. 

In the vacuum chamber, a small granite bench sits on a stiff support structure that runs through vacuum bellows onto a very large, massive, and stiff granite bench. The small granite bench has provisions and a rail system for loading and unloading different test articles. The support structure was designed to have a first frequency $>$ 75 Hz to ensure stability. 

\begin{figure}
\begin{center}
\begin{tabular}{c}
\includegraphics[height=9cm]{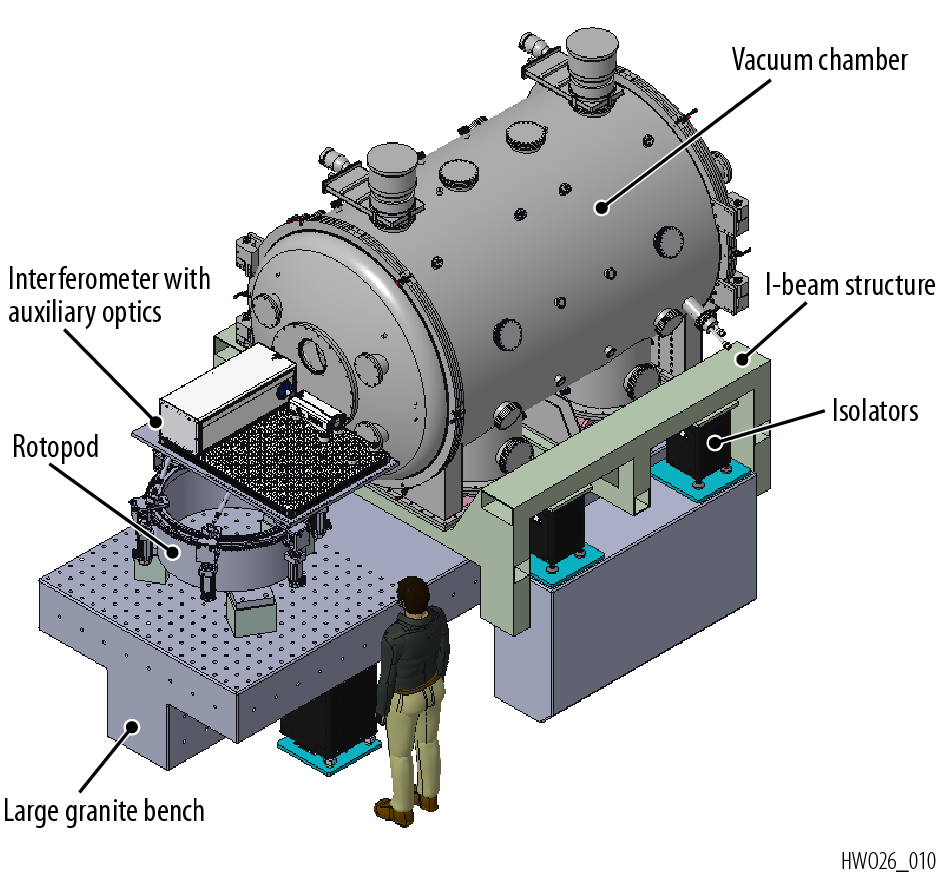}
\end{tabular}
\end{center}
\caption 
{ \label{fig:minimust_zoomout}
Schematic of Mini-MUST, with a 6' human shown for scale.  Mini-MUST can accommodate test articles up to a meter in diameter.} 
\end{figure}

\begin{figure}
\begin{center}
\begin{tabular}{c}
\includegraphics[height=8.5cm]{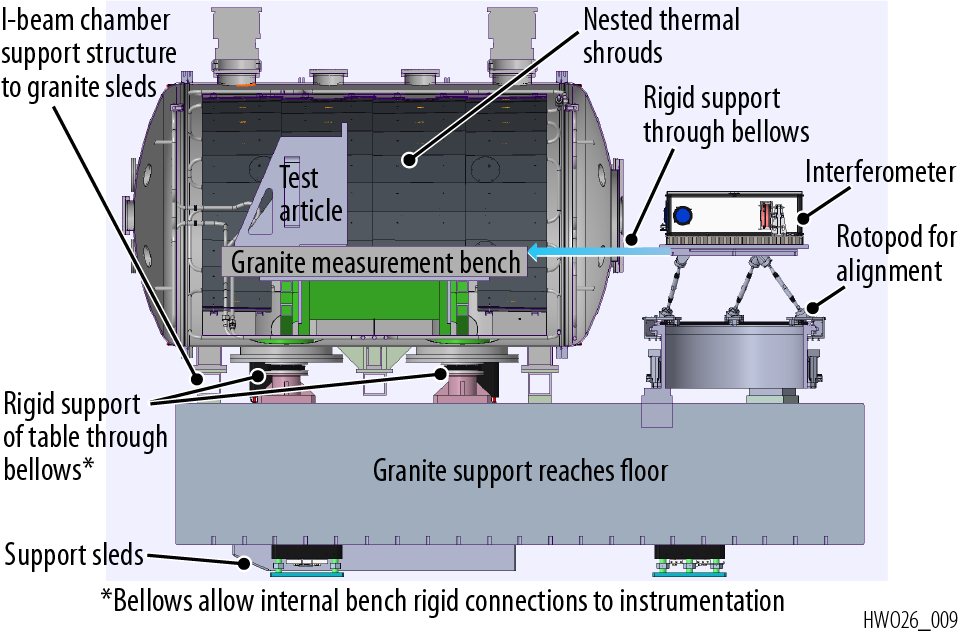}
\end{tabular}
\end{center}
\caption 
{ \label{fig:minimust_cutthrough}
Cross-section of the Mini-MUST chamber and its stable supporting components. Mini-MUST builds upon past experimental set-ups in the \ac{ussl} to enable testing components up to one meter in diameter at picometer levels needed for \ac{trl}5 advancement.} 
\end{figure}

The vacuum chamber itself only contacts the large granite bench and the stiff support structure through the vacuum bellows, which are as flexible as possible. The team consulted with personnel in the NIST Quantum Measurement Division and implemented a vacuum chamber that effectively floats around the measurement bench and is supported by an I-beam structure that is on its own set of isolators on the floor of the \ac{ciaf}. Two schematics of Mini-MUST are shown in Figures~\ref{fig:minimust_zoomout} and \ref{fig:minimust_cutthrough}. 

\begin{figure}
\begin{center}
\begin{tabular}{c}
\includegraphics[height=9cm]{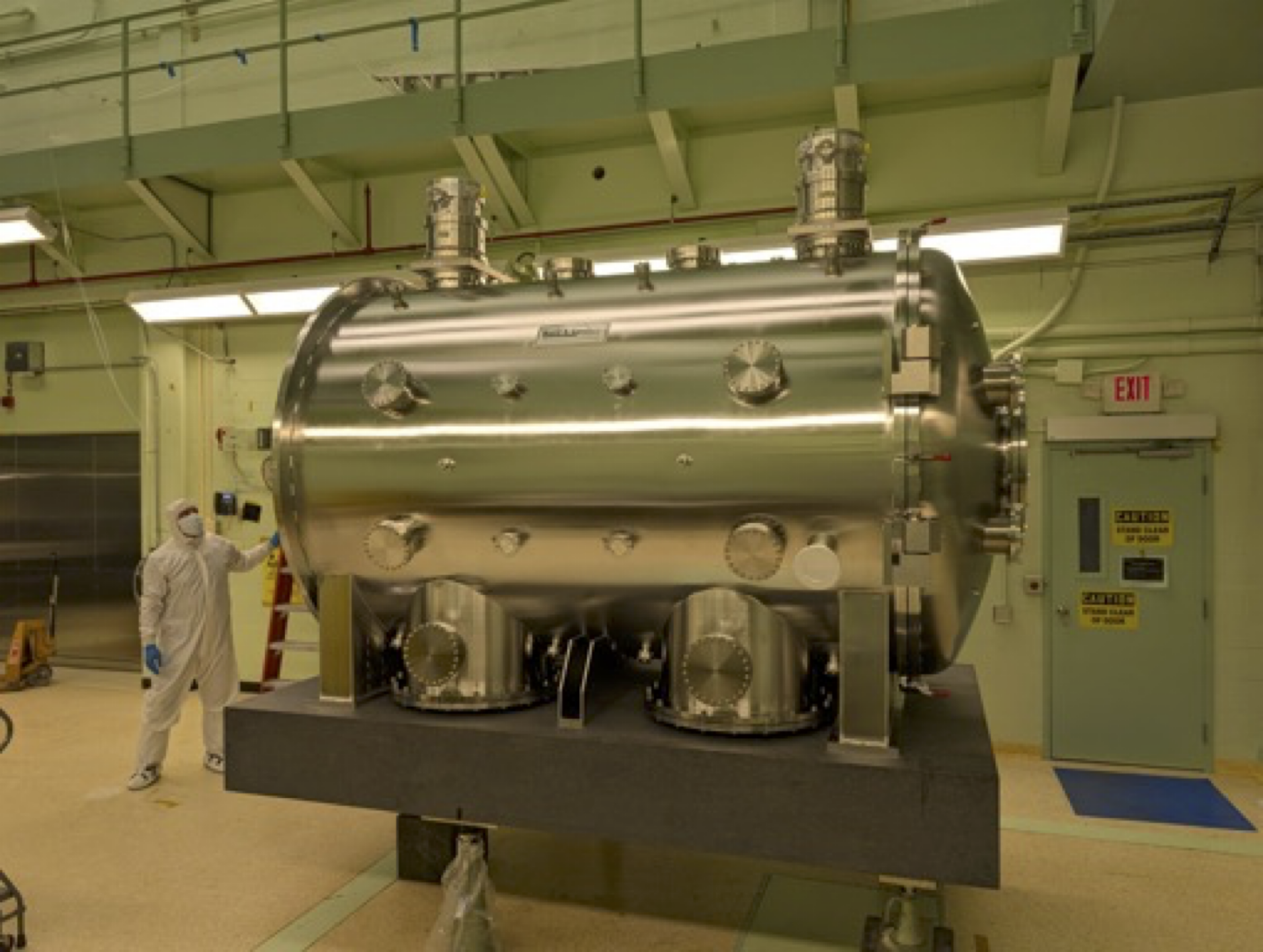}
\end{tabular}
\end{center}
\caption 
{ \label{fig:minimust_CIAF}
Mini-MUST lowered onto an interim granite bench in the CIAF at GSFC. First-light demonstrations are expected in spring 2026.} 
\end{figure}

The thermal vacuum chamber, a critical component of Mini-MUST, was delivered to GSFC in July 2025 and subsequently loaded onto an interim granite bench in the CIAF. Figure~\ref{fig:minimust_CIAF} shows an image of Mini-MUST in the CIAF at GSFC on its concrete pad, denoted by the light green outline on the floor.

Three industry teams (BAE Systems\cite{Cromey2025}, Northrop Grumman\cite{Glassman2025}, and L3Harris) were awarded ROSES D.19~proposals that will bring test articles to the USSL and Mini-MUST to test their stability and performance. The Mini-MUST team is actively working with the teams to deliver models and Mini-MUST specifications for successful testing. 

Detailed information about Mini-MUST can be found in Sitarski et al. (2025)\cite{Sitarski2025}.

\subsection{Coronagraphic Technology Testbeds at JPL}
Proper assessment of \ac{trl} milestones for coronagraphic technologies (coronagraphic mask designs, deformable mirror performance, stabilization techniques, etc.) requires state-of-the-art coronagraphic testbed facilities dedicated to the development and test of \ac{hwo} coronagraphy. \ac{hwo} is leveraging the substantial body of heritage work done at \ac{jpl} in the \ac{hcit} facility run by NASA’s Exoplanet Exploration Program Office, which has produced multiple world record results in high contrast imaging\cite{Seo2019, Allan2023}. Building on shoulders of the HCIT legacy has enabled HWO to move quickly and cost-efficiently towards initial lab demonstrations of the project’s required coronagraph performance.

The coronagraphic testbed strategy for \ac{hwo} at \ac{trl}~5 level consists of two testbeds targeting, first, the quasi-static performance of \ac{hwo} coronagraphy architectures and, second, the capability to stabilize established dark zones over observational timescales and through expected wavefront disturbances. The quasi-static testbed, actively under development at this time, is a renovation of an existing testbed, the \ac{dst2} while the stabilization testbed, funding contingent, will be designed and built completely new for \ac{hwo}.

\subsubsection{DST2 Renovation}

\begin{figure}
\begin{center}
\begin{tabular}{c}
\includegraphics[height=7.3cm]{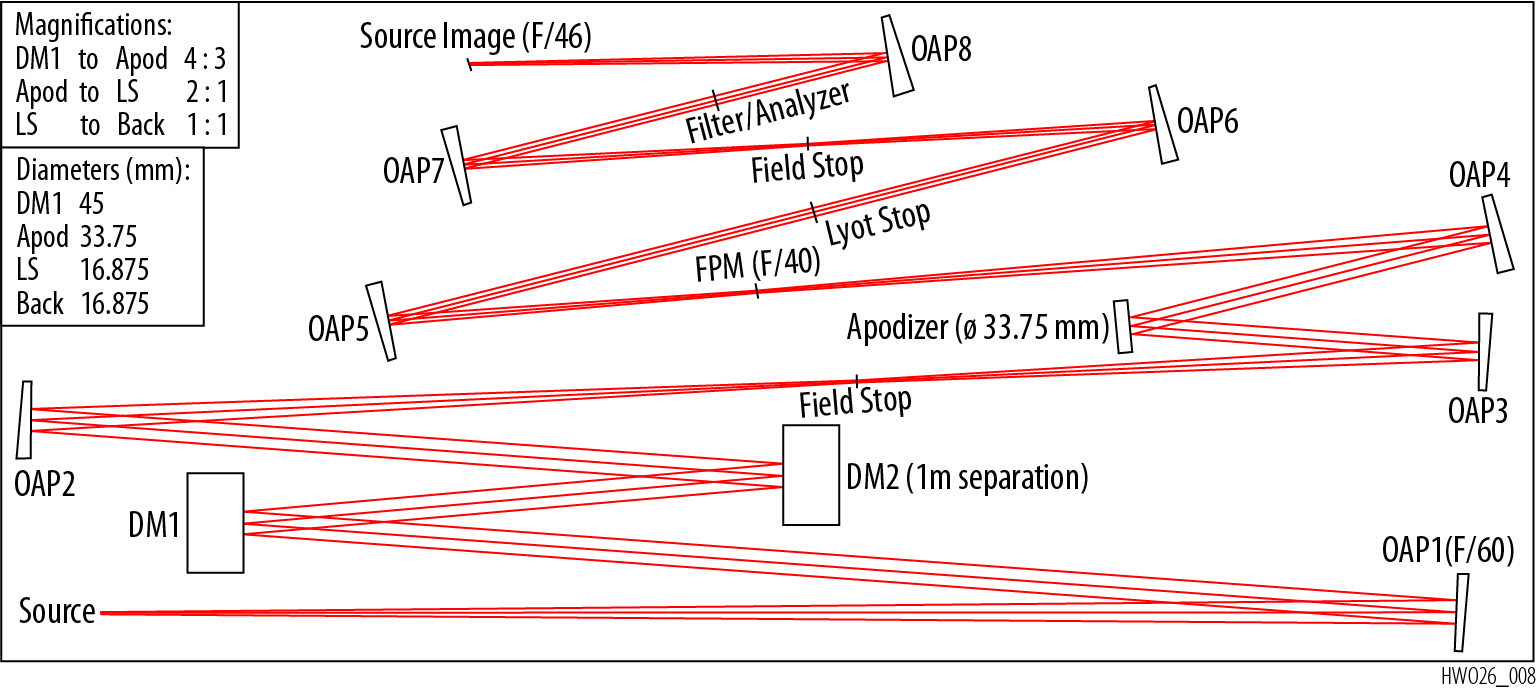}
\end{tabular}
\end{center}
\caption 
{ \label{fig:DST2}
The DST2R primary optical design summary parameters (top left) are shown with the optical layout. The DST2R testbed has apodizer, focal plane, and Lyot-stop planes that can accommodate different coronagraph options under consideration by \ac{htmpo}.} 
\end{figure}

\ac{hwo}'s first coronagraphic testbed targeting technology demonstrations at the TRL5-supporting level is a renovation of \ac{hcit}'s \ac{dst2}, to be operated in the large 12~m vacuum chamber at \ac{jpl}. The redesign (critical path elements shown in Figure~\ref{fig:DST2}) is aimed at providing a coronagraph system that is both capable of HWO performance levels ($\leq$10$^{-10}$ optical contrast ratios) and flexible enough to test all potential coronagraphic architectures of interest to \ac{hwo} scientists and engineers.

This required flexibility has driven optical design choices, especially its 33.75~mm apodizer pupil plane and f/40 beam at the occulting mask focal plane, to accommodate a wide range of starlight nulling technologies such as apodization-type coronagraphs, vortex-type coronagraphs, and hybrid types. The testbed also features a full suite of support functions not shown in Figure~\ref{fig:DST2} such as a polarization analyzer, emission band selector, and focal-plane-rejection Zernike wavefront sensor for pupil-plane wavefront error diagnostics.

The environmental stability of JPL’s 12~m chamber has already proven capable of supporting quasi-static results at 4$\times$10$^{-10}$ using DST1 \cite{Allan2023}. Nevertheless, this renovation of \ac{dst2} also includes improvements to the thermal control system, tackling additional vulnerabilities in its existing thermal control system such as conductive ingress through harnessing and long-time-constant radiative coupling. Initial thermal models predict 1~mK or better stability across all critical elements.

The initial commissioning of this testbed will be done in narrowband and simple Lyot configuration with full model validation before opening the facility to the test of other architectures and broadband configurations at the direction of the HWO project. With the importance of integrated modeling to the observatory as a whole, the testbed and integrated modeling teams are working closely to prepare for and validate testbed models against commissioning results and beyond. A post-commissioning expansion of the testbed is already planned, including an additional pupil relay upstream of the DMs to include injection of simulated amplitude and phase errors from a segmented primary mirror. Inclusion of an integrated field spectrograph contributed by \ac{gsfc} is also planned.

\subsubsection{Exoplanet Imaging Coronagraph for TRL5 and TRL6}

Looking to the future, an entirely new testbed called the \ac{epic5} is planned for handling verification of coronagraph stabilization technologies. With \ac{epic5} slated for a separate, smaller vacuum chamber at \ac{jpl}, this 2-testbed approach will allow faster turn-around on quasi-static results while affording stabilization tests, often interested in timescales on the order of days or weeks, the uninterrupted operation time they require. This approach follows a similar strategy employed by the Roman Coronagraph Instrument to pass several of its early key project milestones for coronagraph performance \cite{Poberezhskiy2014, Trauger2016, Kunjithapatham2015, Shi2016}.

These two testbeds will ultimately be combined in a single testbed capable of demonstrating both quasi-static and stabilization architectures concurrently at TRL6-supporting levels, to form \ac{epic6}. EPIC6 design work will begin as \ac{htmpo} nears and passes \ac{pdr} and is slated for operation in \ac{jpl}'s 12~m vacuum chamber.

\section{Survey of Science Cases Covering Astrophysics and Planetary Science} 
\label{sec:science}

The initial \ac{cml}~3 scientific maturation of the \ac{hwo} was guided by the community-led \ac{start}, operating in close coordination with NASA’s~\ac{tag}. The \ac{start} was co-chaired by Professor Courtney Dressing of the University of California at Berkeley and Chief Scientist John O'Meara at the W.~M. Keck Observatory. The \ac{start} was charged with exploring possible science goals for the mission and providing an initial evaluation of the observatory performance metrics and instrument capabilities required to achieve them. Convened from September 2023 to July 2024, the \ac{start} structured its activities into four thematic working groups encompassing the breadth of potential HWO science: Living Worlds, Solar Systems in Context, Evolution of Elements, and Galaxy Growth. From August 2024 to July 2025, the working groups separately continued their progress after the conclusion of the \ac{start} and \ac{tag} structure, under the guidance of \ac{htmpo}. 

Each working group engaged the broader scientific community in the development of individual \ac{hwo} \ac{scdds}, which established specific science goals and objectives, identified key physical parameters relevant to those science goals, and outlined proposed observational strategies to achieve them. The community developed 73~\ac{scdds} that were provided as input to the \ac{htmpo} and were subsequently compiled into a publicly accessible online archive. Many of these studies were presented at the \ac{hwo25} conference on July 28-31, 2025 and 59 were published in a dedicated special issue of the \textit{Astronomical Society of the Pacific Conference Series}. A comprehensive review of \ac{start}'s activities and principal outcomes will be provided in Dressing et al., in preparation\cite{Dressing.2026}. The following subsections summarize the science working group structure and highlight key findings arising from their collective analyses.

In July 2025, the \ac{csit} was established to build upon the foundation laid by \ac{start}. The \ac{csit} is analyzing science instrument concepts and supporting basic instrument definition, providing advice to \ac{htmpo} in establishing the observatory's primary science objectives and requirements, and serving as key liaisons to the broader scientific community. The \ac{csit} is co-chaired by Professor David Charbonneau of Harvard University and Professor Evgenya Shkolnik of Arizona State University. The \ac{csit} team’s activities are designed to support the \ac{htmpo} through the \ac{mcr}, which will examine the \ac{hwo} architecture concept and whether it can meet its objectives. \ac{htmpo} interfaces with the wider international science community through the \ac{hwo} \ac{sig}, jointly administered by NASA’s three Astrophysics Program Offices: Cosmic Origins, Exoplanet Exploration, and Physics of the Cosmos.

\subsection{Living Worlds}

The Living Worlds working group investigated the detection and characterization of life beyond Earth. This working group was structured around three complementary subgroups that addressed challenges in exoplanet astrobiology: Target Stars and Systems, which built upon existing knowledge to identify optimal stellar targets for HWO observations and assess the capabilities of precursor and contemporaneous observations alongside HWO measurements to constrain critical properties of host stars and their planetary systems; Biosignature Possibilities, which explored the diverse array of potential biosignatures, including biogenic gases, aerosols, surface biosignatures, and technosignatures that could be observed in different planetary conditions; and Biosignature Interpretation, which considered frameworks for assessing potential biosignatures and determined the additional planetary and system-level information necessary to interpret these signals.

The Living Worlds working group both addressed questions focused on detecting Earth-like global biospheres as well as questions of detecting signatures of life ``as we don’t know it'' and prebiotic planets. A major focus was on strategies to detect life during Earth’s evolution over time, with a broader goal to understand the prevalence and diversity of life throughout the galaxy\cite{Arney2025}. This group also investigated the prospects for determining technosignatures elsewhere in the galaxy, including industrial atmospheric pollutants or artificial lights and structures \cite{Kopparapu2025}. The working group considered systematic spectroscopic observations of temperate terrestrial planets using UV/VIS/NIR capabilities spanning 250-1700~nm at varying spectral resolutions. Meaningful constraints on the frequency of planets with life requires observing and characterizing $>$ 25~Exo-Earth candidates over this bandpass, and searching for multiple biosignatures (e.g., oxygen, methane, ozone, surface biosignatures), habitability indicators (e.g., water vapor), and biosignature false positive indicators. This large sample of Earth-like exoplanets would be enabled by a large aperture that has high angular resolution, high sensitivity, and a small coronagraph inner working angle, operating at high contrast, where parallel spectral channels would be beneficial to help manage the long exposure times needed to detect the faint signals. 

\subsection{Solar System in Context}

The \textit{Solar Systems in Context} working group investigated planetary system formation, evolution, and diversity by placing our own solar system within the broader context of exoplanetary systems throughout the galaxy. This working group was organized into four specialized subgroups that span the full range of planetary science investigations: Characterizing Exoplanets, which focused on direct and indirect observations of exoplanets including phase curves, transits, and eclipses; Solar System Observations, dedicated to remote observations of solar system planets, moons, and small bodies; Demographics and Architectures, which considered means to determine occurrence rates and planetary architectures; and Birth and Evolution, which investigated the broader context of planetary system architecture and evolution. 

The \textit{Solar Systems in Context} working group identified dozens of potential science cases for \ac{hwo} ranging from atmospheric characterization of rocky planets to assessments of habitability in solar system ocean worlds \cite{Cartwright2025} and investigations of protoplanetary disk evolution\cite{Ren2025}. For example, the reflectance and polarization on rocky exoplanets could be used to investigate whether they harbor water oceans \cite{LustigYaeger2025}. Planet formation and growth in their environments can be informed by imaging and spectropolarimetry\cite{Berdyugina2025}. One particular investigation proposed a systematic program to determine the time to develop oxygen-rich atmospheres through UV spectroscopy observations at low resolution of approximately 40 Earth-analog planets spanning different stellar ages and environments \cite{Blunt2025}. This working group also considered a comprehensive survey to determine the frequency and compositional diversity of sub-Neptune sized planets to understand planetary demographics\cite{Hu2025}. Closer to Earth, one \ac{scdd} explored how \ac{hwo} could help characterize potentially hazardous asteroids that pose a risk to Earth or human infrastructure throughout the solar system; with its unprecedented sensitivity – better than any other observatories on the ground or space – \ac{hwo} can observe smaller bodies at greater distances than ever before\cite{Dotson2025}. These science cases would be enabled by the high contrast imaging capabilities needed for the \textit{Living Worlds} science cases, UV/VIS/NIR integral field spectroscopy, and near-simultaneous multi-band imaging. Solar system science cases, in particular, place demands on moving object tracking, bright object limits, and field of regard.

\subsection{Galaxy Growth} \label{subsec:GalaxyGrowth}

The \textit{Galaxy Growth} working group investigated HWO-enabled studies of the origins and evolution of galaxies as ecosystems. This working group addressed fundamental questions about how galaxies develop over the history of the observable universe, with particular emphasis on the complex cycles of matter exchange between galaxies and their surrounding environments, including both the circumgalactic and intergalactic media \cite{Tumlinson2017}. The group's organizational structure included four specialized subgroups: Active Galactic Nuclei (AGN) Over Cosmic Time, which investigated AGN and their multi-scale impacts; Ionizing Photons and their History, focused on understanding the stellar sources that fueled cosmic reionization; Intergalactic and Circumgalactic Medium studies; and The Dark Sector, which explored dark matter and dark energy effects through gravitational lensing and other observations.

Numerous topics were studied by the \textit{Galaxy Growth} working group. For example, some \ac{scdd}s considered cosmic reionization by studying galaxies as sources of ionizing radiation for high redshift objects in the rest frame UV\cite{McCandliss2025}. Fundamental astrophysical questions address how early galaxies contributed to the reionization epoch, one of the most significant phase transitions in cosmic history, and continue to shape the ionization state of the intergalactic medium. \ac{hwo} would improve our understanding of the cosmic ecosystem by mapping multiphase gas flow into and away from galaxies\cite{Borthakur2025}. Additionally, the working group's focus on AGN-driven outflows and their multiscale impacts will be used to determine how supermassive black holes regulate star formation and influence their host galaxies' evolution\cite{Zhang2025}. Improved constraints on dark matter can come through measuring the abundance and small-scale power spectrum of low-mass galaxies. This can be achieved by directly imaging dwarf galaxies orbiting Milky Way-analog galaxies \cite{Doppel2025}, or by observing lensed sources behind massive galaxies \cite{He2025}.  This would provide a probe of the underlying cosmological framework that governs large-scale structure formation and could directly constrain the existence of warm dark matter. Many of these science programs would be enabled by a high sensitivity, large collecting area, observatory with high spatial resolution across UV/VIS/NIR and including UV multi-object spectroscopy and integral field spectroscopy.

\subsection{Evolution of the Elements}
\label{subsec:EvolutionElements}

The Evolution of Elements working group developed science cases enabled by HWO and observations of stars and resolved stellar populations across cosmic time.  The rise of elements and molecules is probed through study of the star formation cycle: star formation, the properties of the interstellar medium, stellar evolution, stellar death, and feedback. This working group was organized into three specialized subgroups that addressed complementary aspects of galactic chemical evolution: Star Formation and the Interstellar Medium; Cosmic Explosions, dedicated to understanding supernovae, merger-driven stellar explosions, and gravitational wave source progenitors; and Stellar Populations and Evolution, which conducted detailed spectroscopic and photometric studies ranging from individual Milky Way stars to stellar populations in Local Group galaxies and stellar clusters throughout the universe.

The Evolution of Elements working group posed fundamental questions about how ancient, metal-poor massive stars shaped the early universe's chemical landscape and continue to influence galactic environments. The capabilities of HWO should enable the very first resolved UV spectroscopy of extremely massive, low-metallicity stars ($>$30$M_{\odot}$, $<$0.1$Z_{\odot}$) in nearby analog systems like the IZw18 dwarf starburst galaxy\cite{Senchyna2025}. The working group's investigations into how very massive stars influence their environments across a range of metallicities (0.01-2$Z_{\odot}$) at distances spanning 0.8-15~Mpc provide fundamental observational constraints for models of stellar evolution, feedback, and chemical enrichment in diverse galactic environments\cite{Martins2025}. These studies are complemented by research into the heavy element enrichment history of the universe and investigations of dust extinction curves throughout the Milky Way and Local Group galaxies, which advance our understanding of how stellar nucleosynthesis and explosive phenomena have shaped the chemical complexity of cosmic structures over billions of years. This work will also be used to improve measurements of the Hubble Constant through  measuring Cepheid distances at scales well into the Hubble flow, which may obviate the need for SNe~Ia measurements\cite{Anand2025}. Many of these science programs would be enabled by high spatial resolution across UV/VIS/NIR and including UV multi-object spectroscopy and integral field spectroscopy.

\section{Conclusions}

Significant progress has been made in initiating the \ac{htmpo}, creating a detailed \ac{hwo} Technology Plan, assessing \ac{eac}s 1-3, defining \ac{eac}4 and 5, and progressing the detailed understanding of the system.  At this point, the team has achieved all of the goals set out to achieve \ac{cml}3 as summarized in this paper.  Other collective \ac{cml}3 artifacts held by HTMPO further defines this progress, and activities towards CML4 including definition of \ac{eac}4 and \ac{eac}5 have already commenced.  The team continues to focus on error budgets, concepts of operation, technology specification, scientific definition, and is ready to make major investments in technology.  Overall, the architecture has converged to be a 6-8~m off-axis telescope, whose primary mirror is segmented, that is housed in a serviceable, room-temperature observatory. Over the coming two years, \ac{eac}4 and 5 will be studied and traded to enable a final single baseline concept at \ac{mcr}.

\section{Code, Data, and Materials Availability}
\label{sect: code}
Data and analysis code can be provided upon reasonable request.

\section*{Disclosures}
The authors declare that there are no financial interests, commercial affiliations, or other potential conflicts of interest that could have influenced the objectivity of this research or the writing of this paper.

\subsection* {Acknowledgments}
The authors would like to acknowledge the contributions of engineering and science teams at NASA Centers during this time of change and uncertainty, including NASA's Goddard Space Flight Center, Jet Propulsion Laboratory, Ames Research Center, and Marshall Space Flight Center. The team acknowledges the contributions from the \ac{start} committee and the international science community for creating visionary science cases, and the \ac{csit} for their contributions into establishing representative science programs and technical capabilities for \ac{hwo}. Technology recommendations and architectural concepts were put forward by numerous NASA, industry, and academic members. Each member has made an important contribution to the \ac{hwo} mission.


\bibliography{report}   
\bibliographystyle{spiejour}   


\vspace{2ex}\noindent\textbf{Lee Feinberg} was the Principal Architect for the \ac{hwo}. Her received his BS and MS degrees in optics and applied physics in 1987 and 1995, respectively. He is a Goddard Space Flight Center Senior Fellow, an SPIE Senior Fellow, and has won the Presidential Rank award for his more than 20 years of leading \ac{jwst}. 

\vspace{2ex}\noindent\textbf{Breann Sitarski} is the deputy principal architect, Goddard Space Flight Center Testbeds lead, and optical systems engineer for \ac{hwo}. Breann’s research interests are in precision metrology, optical alignment and testing, wavefront sensing and control for large telescopes, and precision astrometry. She earned her Ph.D. in astrophysics at the University of California, Los Angeles under the guidance of Professors Andrea Ghez and Michael Fitzgerald. She was previously the deputy project systems engineer, optical systems scientist, and alignment lead at the Giant Magellan Telescope.

\vspace{2ex}\noindent\textbf{Michael McElwain} is the \ac{hwo}’s Observatory Scientist, \ac{jwst}’s Observatory Project Scientist, and the Exoplanets and Stellar Astrophysics Laboratory Chief at NASA’s Goddard Space Flight Center. Michael’s research interests include large space telescopes and specialized instrumentation for exoplanet characterization. He earned his PhD at the University of California, Los Angeles, where he was a member of the Infrared Laboratory.  He was a postdoctoral researcher at Princeton University and policy fellow at the National Academies before joining NASA.

\vspace{2ex}\noindent\textbf{Giada Arney} is the \ac{hwo}’s Project Scientist at NASA’s Goddard Space Flight Center. Giada’s research interests include the search for life on exoplanets, planetary habitability, Early Earth as an analog for Earth-like exoplanets, and Venus as a possibly once-habitable planet. She earned her dual-title PhD in astronomy and astrobiology at the University of Washington in Seattle.  She joined NASA Goddard Space Flight center as a NASA postdoctoral program research fellow prior to becoming a civil servant.

\vspace{2ex}\noindent\textbf{Caleb Baker} 

\vspace{2ex}\noindent\textbf{Matthew D. Bolcar} is an optical systems engineer at NASA’s Goddard Space Flight Center, where he is a member of the Wavefront Sensing and Control group in the Optics Branch. He currently serves as the optical systems lead for the \ac{ngrst} and the chief technologist for the Habitable Worlds Observatory Technology Maturation Project Office. Additional projects that he has worked on include the Large Ultraviolet/Optical/Infrared Surveyor (LUVOIR) concept study, the Thermal Infrared Sensor (TIRS) instrument on LandSat 8, the Advanced Topographical Laser Altimeter System (ATLAS) instrument on ICESat 2, the Visible Nulling Coronagraph, and the Wide-field Imaging Interferometer Testbed. Prior to coming to Goddard, he received a B.S. in Engineering Physics from Cornell University in 2002 and earned his Ph.D. under Professor James R. Fienup at the University of Rochester in 2009.

\vspace{2ex}\noindent\textbf{Marie Levine} received a Ph.D. from the California Institute of Technology and has been working at the Jet Propulsion Laboratory for 35 years as an expert in precision optical systems and integrated modeling. She is currently the HWO Integrated Modeling lead within the NASA Mission Systems Engineering team. She held a similar position for nearly 14 years on the James Webb Space Telescope. Prior to that Marie was the Technology Manager of the Exoplanet Exploration Program, the Systems Manager for the Terrestrial Planet Finder Coronagraph, and the PI on two flight experiments IPEX-I and II for the on-orbit characterization of microdynamics in precision structures. Marie is also a member of the NASA Standing Review Board for the Roman Space Telescope. She has received many awards including the NASA Exceptional Engineering Medal, the NASA Exceptional Service Medal and was designated Woman of the Year by US Representative Adam Schiff where here accomplishments were read in front of Congress.

\vspace{2ex}\noindent\textbf{Alice (Kuo-Chia) Liu} serves as HWO's Mission Systems Engineer and the Integrated Modeling and Error Budget Lead for the Nancy Grace Roman Space Telescope. She has worked at NASA Goddard Space Flight Center for over 23 years, supporting various flight missions and holding multiple leadership roles. She earned her PhD from the Massachusetts Institute of Technology in Dynamics and Control, focusing on optimizing staged control systems for large space interferometers. Her doctoral research was supported by a Michelson Fellowship. She began her career as the Jitter Lead for the Solar Dynamics Observatory and the Attitude Control System Lead for the Terrestrial Planet Finder-Coronagraph study.

\vspace{2ex}\noindent\textbf{Bertrand Mennesson} 

\vspace{2ex}\noindent\textbf{Aki Roberge} is the \ac{hwo}’s Pre-Formulation Scientist at NASA’s Goddard Space Flight Center. Aki's research interests include multi-wavelength studies of exoplanet formation, UV astrophysics, and space observatory concept development. She earned her PhD in astrophysics at the Johns Hopkins University in Baltimore MD. She held postdoctoral research fellowships at the Carnegie Institution for Science and NASA Goddard Space Flight Center prior to joining the latter as a civil servant scientist. 

\vspace{2ex}\noindent\textbf{J. Scott Smith} is the \ac{hwo} Project Manager and the \ac{ngrst} Telescope Manager.  Scott was the Roman telescope manager from Phase A through the delivery of the telescope.  Scott supported JWST during the TRL-6 demonstrations through telescope and instrument level testing in the areas of wavefront sensing and optics. Scott earned his masters degree from the University of Maryland.

\vspace{2ex}\noindent\textbf{Feng Zhao} 

\vspace{2ex}\noindent\textbf{John Ziemer} is a Principal Engineer at NASA’s Jet Propulsion Laboratory, where he manages strategic mission and technology development for the Space and Earth Science Formulation Office. He currently serves as the Pre-Formulation Architect for NASA’s Habitable Worlds Observatory and Program Manager for the PRIMA far-infrared probe mission concept. John earned his Ph.D. and M.A. from Princeton University and his B.S.E. from the University of Michigan, specializing in plasma science and electric propulsion. His work at JPL focuses on integrating advanced technologies with compelling science to define and enable next-generation space missions.

\vspace{1ex}

\listoffigures
\listoftables

{\noindent \bf ACRONYMS}
\begin{small}
\begin{acronym}[parsep=0pt]
\acro{astro2020}[Astro2020]{Pathways to Discovery in Astronomy and Astrophysics for the 2020s}
\acro{acs}[ACS]{Attitude Control System}
\acro{adi}[ADI]{Angular Differential Imaging}
\acro{assist}[ASSIST]{Active Segmented Surrogate for Integrated System Tests}
\acro{cad}[CAD]{Computer-Aided Design}
\acro{ciaf}[CIAF]{Calibration, Integration, and Alignment Facility}
\acro{cml}[CML]{Concept Maturity Level}
\acro{conops}[ConOps]{Concept of Operations}
\acro{csit}[CSIT]{Community Science and Instrument Team}
\acro{dst2}[DST2]{Decadal Survey Testbed 2}
\acro{eac}[EAC]{Exploratory Analytic Case}
\acro{ee}[EE]{Encircled Energy}
\acro{epic5}[EPIC5]{Exoplanet Imaging Coronagraph for TRL5}
\acro{epic6}[EPIC6]{Exoplanet Imaging Coronagraph for TRL6}
\acro{for}[FOR]{Field of Regard}
\acro{frn}[FRN]{Flux Ratio Noise}
\acro{fuv}[FUV]{far ultraviolet}
\acro{gomap}[GOMAP]{Great Observatory Mission and Technology Maturation Program}
\acro{gsfc}[GSFC]{Goddard Space Flight Center}
\acro{hcit}[HCIT]{High Contrast Imaging Testbed}
\acro{hicat}[HiCAT]{High-Contrast imager for Complex Aperture Telescopes}
\acro{hri}[HRI]{High-Resolution Imager}
\acro{host}[HOST]{Habitable Worlds Observatory Systems Testbed}
\acro{htmpo}[HTMPO]{HWO Technology Maturation Project Office}
\acro{hwo}[HWO]{\textit{Habitable Worlds Observatory}}
\acro{hwo25}[HWO25]{\textit{Towards the Habitable Worlds Observatory: Visionary Science and Transformational Technology}} 
\acro{im}[IM]{Integrated Modeling}
\acro{jpl}[JPL]{Jet Propulsion Laboratory}
\acro{jwst}[JWST]{\textit{James Webb Space Telescope}}
\acro{los}[LOS]{line-of-sight}
\acro{mems}[MEMS]{micro-electrical-mechanical system}
\acro{mcr}[MCR]{Mission Concept Review}
\acro{ngrst}[NGRST]{\textit{Nancy Grace Roman Space Telescope}}
\acro{nuv}[NUV]{near ultraviolet}
\acro{os}[OS]{Observing Scenario}
\acro{pdr}[PDR]{Preliminary Design Review}
\acro{psf}[PSF]{point spread function}
\acro{rms}[RMS]{Root Mean Square}
\acro{scdd}[SCDD]{Science Case Development Document}
\acro{scdds}[SCDDs]{Science Case Development Documents}
\acro{sig}[SIG]{Science Interest Group}
\acro{sls}[SLS]{Space Launch System}
\acro{start}[START]{Science, Technology, and Architecture Review Team}
\acro{stop}[STOP]{Structural, Thermal, and Optical Performance}
\acro{stsci}[STScI]{Space Telescope Science Institute}
\acro{tag}[TAG]{Technical Assessment Group}
\acro{trl}[TRL]{Technology Readiness Level}
\acro{ule}[ULE]{Ultra-Low Expansion}
\acro{ussl}[USSL]{Ultra-Stable Structures Lab}
\acro{vpm}[VPM]{Vertical Primary Mirror}
\acro{wfe}[WFE]{wavefront error}
\acro{wfsc}[WFS\&C]{wavefront sensing and control}
\end{acronym}
\end{small}

\end{spacing}
\end{document}